# *ANALYSIS OF FINANCIAL CREDIT RISK USING MACHINE LEARNING*

**Jacky C. K. Chow**

Aston Business School

**Aston University**

**Birmingham, United Kingdom**

**This dissertation is submitted for the degree of Master of Business Administration**

**April 2017**

*To my courageous, wise, and loving mother*

*Thank you for your patience and time in teaching and taking care of me*

*Thank you for giving up everything for me*

*Thank you for all your prayers*

*Your unconditional love and encouragements made me who I am*

*I will miss you forever, rest in peace*

# DECLARATION

I declare that I have personally prepared this report and that it has not in whole or in part been submitted for any other degree or qualification. Nor has it appeared in whole or in part in any textbook, journal or any other document previously published or produced for any purpose. The work described here is my own, carried out personally unless otherwise stated. All sources of information, including quotations, are acknowledged by means of reference, both in the final reference section, and at the point where they occur in the text.

# ABSTRACT

Corporate insolvency can have a devastating effect on the economy. With an increasing number of companies making expansion overseas to capitalize on foreign resources, a multinational corporate bankruptcy can disrupt the world's financial ecosystem. Corporations do not fail instantaneously; objective measures and rigorous analysis of qualitative (e.g. brand) and quantitative (e.g. econometric factors) data can help identify a company's financial risk. Gathering and storage of data about a corporation has become less difficult with recent advancements in communication and information technologies. The remaining challenge lies in mining relevant information about a company's health hidden under the vast amounts of data, and using it to forecast insolvency so that managers and stakeholders have time to react. In recent years, machine learning has become a popular field in big data analytics because of its success in learning complicated models. Methods such as support vector machines, adaptive boosting, artificial neural networks, and Gaussian processes can be used for recognizing patterns in the data (with a high degree of accuracy) that may not be apparent to human analysts. This thesis studied corporate bankruptcy of manufacturing companies in Korea and Poland using experts' opinions and financial measures, respectively. Using publicly available datasets, several machine learning methods were applied to learn the relationship between the company's current state and its fate in the near future. Results showed that predictions with accuracy greater than 95% were achievable using any machine learning technique when informative features like experts' assessment were used. However, when using purely financial factors to predict whether or not a company will go bankrupt, the correlation is not as strong. More features are required to better describe the data, but this results in a higher dimensional problem where the thousands of published companies' data are insufficient to populate this space with high enough density. Due to this "curse of dimensionality", flexible nonlinear models tend to over-fit to the training samples and thus fail to generalize to unseen data. For the high-dimensional Polish bankruptcy dataset, simpler models such as logistic regression could forecast a company's bankruptcy one year into the future with 66.4% accuracy.

# ACKNOWLEDGEMENTS

Thank you, Alison, Angela, Bahar, Brigitte, Daniel, Hiro, Karen, Kirit, Malcolm, Matt, Nathan, and Uwe for everything you have taught me about business management.

Thank you Adekola, Conor, Farhan, and Omar for all the thought provoking discussions. I enjoyed working with all of you over the years.

Thank you Henk and Jeroen for supporting my endeavour of broadening my knowledge by pursuing a business degree.

Thank you Alex, Angelo, Andrea, Arun, Daniela, Elisa, Elise, Flavia, Giovanni, Ignazio, Jan, Job, Laura, Laurens, Luca, Matteo, Niek, and Victoria for making my time in Europe less lonely and full of joy.

Thank you Ajeesh, Bailey, Bryan, Colin, Jason, Karen, Kate, Kathie, Kris, Kristi, Louis, Mark, Matt, Michael, Michelle, Monica, Trista, Suzanne, and Terra for making me feel positive and look forward to being at work every day.

Thank you K'dee for proof-reading this thesis and being supportive all the time.

Thank you Marie Skłodowska-Curie Actions for funding this research.

The data analysed in this thesis has been provided by the University of California, Irvine. Lichman, M. (2013). UCI Machine Learning Repository [http://archive.ics.uci.edu/ml]. Irvine, CA: University of California, School of Information and Computer Science.

# CONTENTS

# LIST OF TABLES

# List of Figures

# List of Abbreviations and Acronyms

| | |
|---|---|
| AdaBoost | Adaptive Boosting |
| ANN | Artificial Neural Network |
| AUC | Area Under Curve |
| Bagging | Bootstrap Aggregating |
| BLUE | Best Linear Unbiased Estimate |
| CV | Cross-Validation |
| GP | Gaussian Processes |
| ISOMAP | Isometric Feature Mapping |
| K-D Tree | K-Dimensional Tree |
| LDA | Linear Discriminate Analysis |
| MLE | Maximum Likelihood Estimate |
| PCA | Principal Component Analysis |
| ROC | Receiver Operating Characteristic |
| SVM | Support Vector Machine |

# LIST OF APPENDICES

# 1 INTRODUCTION

This is the century of data. Harvard Business Review recently published an article which named Data Scientist the "sexiest job" of the 21st century (Davenport & Patil, 2012). With big data about consumers, marketing, operations, accounting, economics, etc. already widely available to most corporations, the last piece of the puzzle appears to be extracting valuable information that can be interpreted by humans, using techniques such as data-mining or machine learning. Corporations are already restructuring their company strategies to reap the benefits from machine learning. A survey done by the Accenture Institute for High Performance indicated that more than 40% of large corporations are already using machine learning to boost their marketing and they can attribute approximately 38% of their sales improvement to machine learning. In addition, 76% of these corporations believe that machine learning will be a key component of their future sales growth (James Wilson, Mulani, & Alter, 2016).

The applications of machine learning to business are broad; aside from targeted sales and market segmentation, it can be used for inventory optimization based on demand forecasting, personalized customer service and customer segmentation, and many more (Chen, Chiang, & Storey, 2012). The domain that will be studied extensively in this thesis is financial credit risk assessment. Terminology such as credit rating/scoring, bankruptcy prediction, and corporate financial distress forecast will be used interchangeably and together they will be referred to as "financial credit risk assessment" (Chen, Ribeiro, & Chen, 2016). The reason for such simplification is that (from a probabilistic machine learning perspective) all these problems can be cast into a binary classification problem in the final stage, e.g. Will this company be bankrupt by next quarter? Answer: Yes or No.

Bankruptcy prediction dates back more than two centuries where most assessments were done qualitatively (Bellovary, Giacomino, & Akers, 2007; Li & Miu, 2010; de Andrés, Landajo, & Lorca, 2012). It was not until the 20th century that more quantitative (and less subjective) techniques became popular; some examples include the seminal univariate analysis work of Beaver (Beaver, 1966) and multiple discriminant analysis work of Altman in the 1960s (Altman, 1968). Their work demonstrated the ability to predict a company's failure up to five years in advance. Such information is an asset not only to creditors, auditors, stockholders, senior management, etc. because it can have a direct effect on them, but also to many other stakeholders such as suppliers and employees (Wilson & Sharda, 1994).

To understand the significance and possible impacts of corporate bankruptcy on the rest of society, it is worthwhile to revisit the largest bankruptcy in world history, Lehman Brothers Holdings Inc. Caused by social irresponsibility in management and triggered by their exposure to the subprime mortgage crisis in the United States, on September 15, 2008, the fourth largest investment bank in the United States declared bankruptcy (Williams, 2010). The global economy went from bad to worse. Almost six million jobs were lost (the U.S. unemployment rate doubled), Dow Jones industrial average dropped 5000 points, and an estimated $14 trillion of wealth was destroyed (Shell, 2009). On the same day that Lehman Brothers declared bankruptcy, The European Central Bank and The Bank of England in London injected more than $50 billion into the market to calm the world economy (Ellis, 2008). American Broadcasting Company (ABC) News described it as a "financial tsunami" and even compared it to the Great Depression in the 1930s. To many people, this event may have seemed sudden, but such financial disaster did not happen overnight; there were patterns in the data months (even years) prior to this incident that most people failed to recognize (Demyanyk & Hasan, 2010). A reliable financial distress forecasting system could have identified financial issues and challenges prior to the actual bankruptcy. Such a system would be beneficial to companies in various industries worldwide, as company failures are certainly not exclusive to the American economy.

As stated in a recent business article from Forbes: "Machine learning is redefining the enterprise in 2016" (Columbus, 2016). Therefore, it is critical for business managers, banks, investors, and other stakeholders to develop an understanding and intuition about how these algorithms can be beneficial to their decision-making process. This thesis will investigate the value and limitations of machine learning techniques for businesses, with a focus on financial credit risk assessment. Popular machine learning techniques such as logistic regression, support vector machines, decision trees, AdaBoost, artificial neural networks, and Gaussian processes will be explored. The efficacy of such tools for expressing business well-being and improving business performance will be analysed.

# 2 BACKGROUND

## 2.1 Corporate Bankruptcy

Between the years 2012 and 2016, an average of 32 176 bankruptcies have been filed annually by businesses in the U.S. alone (United States Courts, 2016). In the European Union, there are approximately 200 000 corporate bankruptcy filings every year (Mańko, 2013). Some scholars have argued that this is a natural process of a free-market economy (White, 1989). As competition arises, companies will be forced to optimize their operations to increase efficiency and maximize their output given their resources. This will continue until a market equilibrium is achieved where supply and demand are balanced. During this process, companies that are inferior compared to their competitors will be eliminated and bankruptcy can be seen as a filter to remove economic inefficiencies from the market.

In the U.S., most corporations will file for bankruptcy under either Chapter 7 (liquidation) or Chapter 11 (reorganization) in Title 11 of the United States Bankruptcy Code. In the case of liquidation, a trustee will be appointed to sell all the company's assets in order to pay off the debt. Since the total liabilities will usually exceed the total assets at this stage of the business, debt holders will be paid back based on the absolute priority rule. Essentially, the most risk averse investors, who agreed to the least amount of monetary return even during the peak performance periods of the company will be repaid first, and this will go down different tiers of investors until the money runs out. For example, the secured creditors are paid before shareholders (Newton, 2003). Most companies will only file for liquidation as a last resort; if possible, companies would prefer to restructure their business's assets and debts under Chapter 11. Although this allows the business owners to continue operating their business, a committee will be working closely with the owners and managers to turn the business around. Some business decisions can only be made if the court believes it is in the interest of the creditors. By this point it is difficult for a company to become profitable once again, but there have been exceptions (e.g. General Motors and United Airlines). Also, reorganization is not always possible. Companies need to be able to demonstrate that their assets are worth significantly more if they stay in operation than if they were simply sold.

In today's global economy, it is common for businesses to operate across geographic boundaries. Corporate bankruptcy becomes a lot more complicated when a company is operating in multiple countries under different jurisdictions. The brief overview of corporate bankruptcy described above is specific to the United States. The legal system around bankruptcy can vary wildly between countries and this can result in duplicate prosecution claims, increased administrative cost, race to file, inefficiency in capital allocation, and many more complications if not administered properly (Guzman, 1999). Scholars have continued to debate between universalism and territorialism when it comes to insolvency of international businesses (Tung, 2001). Under the universalism perspective, the bankruptcy should follow the law of the business's home country and all the creditors should be paid accordingly without filing numerous claims. However, this is mostly an idealistic policy. In reality, countries are typically unwilling to transfer capital over to another country's economy at the expense of their domestic investors. Not to mention it is challenging to enforce foreign law in domestic disputes. What has emerged is the territorial regime where everyone scavenges for themselves. There have been active collaborations between some countries to address this issue of cross-border insolvency; for instance, the Nordic Bankruptcy Convention governs international corporate bankruptcy in Scandinavia.

## 2.2 Machine Learning Models for Financial Distress Prediction

An unbiased objective prediction of a company's probability of going bankrupt can be a useful management tool. Numerous methods have been proposed for bankruptcy prediction. Some review papers have attempted to categorize them into statistical methods, intelligent systems, data mining, and machine learning techniques. However, the boundaries between these disciplines are slowly vanishing; statistical methods like logistic regression and intelligent systems such as support vector machines are now taught in almost every machine learning course. Therefore, all these data-driven learning methods for continuous and discrete outputs will simply be considered as machine learning techniques. In general, the machine learning pipeline consists of pre-processing (e.g. standardization and centroid reduction), dimensionality reduction (e.g. principal component analysis), training (e.g. learning parameters), model-selection (e.g. validation), and testing (e.g. accuracy assessment of the predictions) (Murphy, 2012), as illustrated in Figure 1.

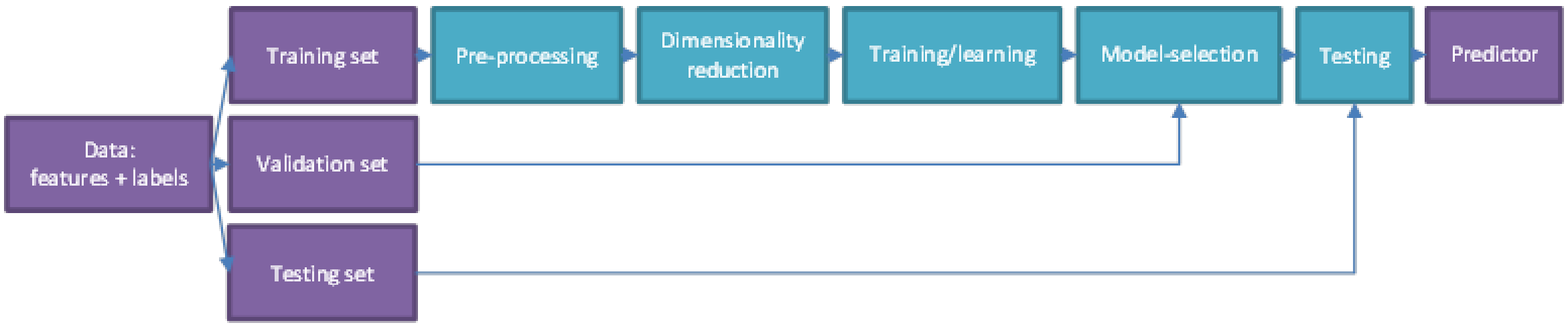

*Figure 1: Overview of a typical machine learning pipeline*

Before looking at the state-of-the-art methods in this field, it is helpful to look at the developments and progression in the past. Qi (2013) presented a concise history of machine learning for corporate bankruptcy prediction, highlighting some major research initiatives in the past 50 years. Since the work of Beaver and Altman in the 1960s, an increasing interest in quantitative assessment of financial credit risk can be observed. In the 1970s, methods such as ordinary least squares regression (Meyer & Pifer, 1970), discriminant analysis (Deakin, 1972), and logistic regression (Martin, 1977) were deployed for bankruptcy classification tasks. Variations of discriminant analysis also soared above univariate analysis in terms of performance due to its ability to account for correlation between variables. During this time, Altman was already able to achieve classification accuracy above 95% one period before bankruptcy and above 70% three periods before bankruptcy (Haldeman, Altman, & Narayanan, 1977). By the 1980s, logit analysis (Ohlson, 1980), factor analysis (West, 1985), and other similar methods were introduced to the field of bankruptcy prediction. In the 1990s, the community saw

Altman's original Z-score method being extended to private firms, non-manufacturers, and firms in emerging markets (Altman, 1995). On one hand this indicated the applicability of the same machine learning method to different markets, but on the other hand it highlighted the fact that the probability distributions coming from different markets are unique, making it difficult to generalize the training results beyond the market being studied. For example, applying an accurate prediction model trained on American steel industry business data to the European apparel market will likely result in high classification error.

By the 2000s, Bayesian methods had been adopted to combine financial ratios and maturity schedule factors (Philosophov, Batten, & Philosophov, 2007). In the 2010s, features beyond financial ratios (such as accounting-based measures, equity prices, firm characteristics, industry expectations, macroeconomic indicators, and agents' opinions) were used for prediction (Altman, Fargher, & Kalotay, 2011; Li, Lee, Zhou, & Sun, 2011). The growing number of features being used in the model was likely a result of the limited expressiveness of financial ratios on the complex system of financial distress. Measuring financial ratios is not equivalent to observing "real market characteristics"; these hidden variables need to be inferred from indirect measurements. Including more features typically paints a better picture of the actual business situation, but it does come at a price. Not only can it be expensive and cumbersome to collect that information (if not impossible due to confidentiality), it turns bankruptcy prediction into a high-dimensional classification problem. This increases the complexity of utilizing machine learning methods effectively as they need to combat the curse of dimensionality. The number of features under consideration in relevant articles usually ranges from 1 to 57 (du Jardin, 2009). With these challenges, novel solutions have also surfaced in recent years. These include various modifications of artificial neural networks such as probabilistic neural network (Yang, Platt, & Platt, 1999) and self-organizing maps (Kaski, Sinkkonen, & Peltonen, 2001). Other modern day alternative solutions include decision trees (Li, Sun, & Wu, 2010), case-based reasoning (Li & Sun, 2011), support vector machines (Trustorff, Konrad, & Leker, 2011), soft computing (Demyanyk & Hasan, 2010), genetic algorithms (Davalos, Gritta, & Adrangi, 2007), AdaBoost (Sun, Jia, & Li, 2011), and Gaussian process (Peña, Martínez, & Abudu, 2011).

It is not the intention of this thesis to review all state-of-the-art machine learning methods. Instead, the fundamental principles of a subset of popular modern day machine learning methods relevant to bankruptcy prediction will be explained in the following subsections.

### 2.2.1 Linear Regression and Logistic Regression

Least-squares regression/estimation/adjustment is one of the most popular mathematical tools in statistics, engineering, and econometrics. Given a set of N observations $\vec{y}$ (a.k.a. the dependent variable or regressand) and some feature map $\vec{\Phi}$ of explanatory variables $\vec{x}$ (a.k.a. regressor), the best set of weights, $\vec{w}$, that minimizes the square residual errors (i.e. L2-norm) can be calculated. The objective function for such linear models can be expressed mathematically by Equation 1. It can further be shown that this gives the Best Linear Unbiased Estimate (BLUE) of the unknown weights (Förstner & Wrobel, 2016). If the residuals, $\vec{e}$, follow a Gaussian probability distribution, then the least-squares solution can be proved to coincide with the Maximum Likelihood Estimate (MLE) (Bishop, 2006).

$$\min_{w} \vec{e}^{\,T} \vec{e} = \min_{w} \left(\vec{y} - \vec{\Phi}(\vec{x})\vec{w}\right)^{T} \left(\vec{y} - \vec{\Phi}(\vec{x})\vec{w}\right) \qquad 1$$

The choice of a linear model is generally not prohibitive for two reasons: first, many important business problems can be expressed using a linear relationship after some feature mapping, and second, complex nonlinear models can be linearized using a first-order Taylor series expansion. For instance, a linear model can be used to answer questions like what is the expected price of detached houses in Calgary (i.e. $\vec{y}$) given data such as age of house, lot size, number of bedrooms, and number of bathrooms (i.e. $\vec{x}$)? Given a set of training data N that is greater than or equal to the number of unknowns M, the optimal weights in Equation 1 can be estimated using Equation 2.

$$\vec{w} = \left(\vec{\Phi}(\vec{x})^{T} \vec{\Phi}(\vec{x})\right)^{-1} \vec{\Phi}(\vec{x})\, \vec{y} \qquad 2$$

If a complex nonlinear functional model is used, the problem can still be linearized locally using a first-order Taylor series expansion. The only difference is that Equation 2 will need to be applied multiple times, starting at some good initial approximation of the weights. This yields what is commonly known as the Gauss-Newton update in an iterative least-squares scheme for solving nonlinear problems (Nocedal & Wright, 2006).

Linear regression takes continuous inputs (e.g. recent years' Du Pont Ratio, profit margin, efficiency ratio, acid-test ratio, cash ratio, debt ratio, earnings per share, etc.) and outputs a continuous variable (e.g. next year's expected profit). This can already be useful for forecasting a company's financial distress. If the predicted profit is a large negative number and there is not enough cash flow to cover the deficit, it is likely that the company will go bankrupt. However, this requires an intelligent and experienced agent to decide what that threshold is for a particular company; i.e. how low does the predicted profit need to be to declare a financial crisis? Instead of a human expert, a computer can be trained to set that threshold automatically by supplying the artificial agent with many binary labels (e.g. 0 if the company did not file for bankruptcy and 1 if the company did) as the dependent variable $\vec{y}$. In this case, the linear regression model needs to be modified to accommodate for his discrete output.

Previously, $\vec{y}$ was assumed to follow a Gaussian distribution; but now since it can only take on a value of zero or one, it follows a Bernoulli distribution instead. Furthermore, the linear combination of the input variables needs to be restricted between zero and one hundred percent to make it a probability, which can be achieved using a sigmoid/logit function (Equation 3); the shape of a sigmoid function is shown in Figure 2. After these modifications, this mathematical formulation is often known as logistic regression or logit regression, Equation 4 (Murphy, 2012). This is an example of a discrete choice model in economics (Train, 2004). When used to make a prediction, the output is now binary: in the near future, either the company will declare bankruptcy or not. This makes the interpretation of the results simple enough that even non-experts in finances (e.g. human resources managers and IT managers) can easily understand. Such a financial distress forecast is more than just a bankruptcy predictor for investors. It can also be used by managers to monitor the health of their business. This feedback can help steer business strategies away from insolvency and increase the probability of profitability.

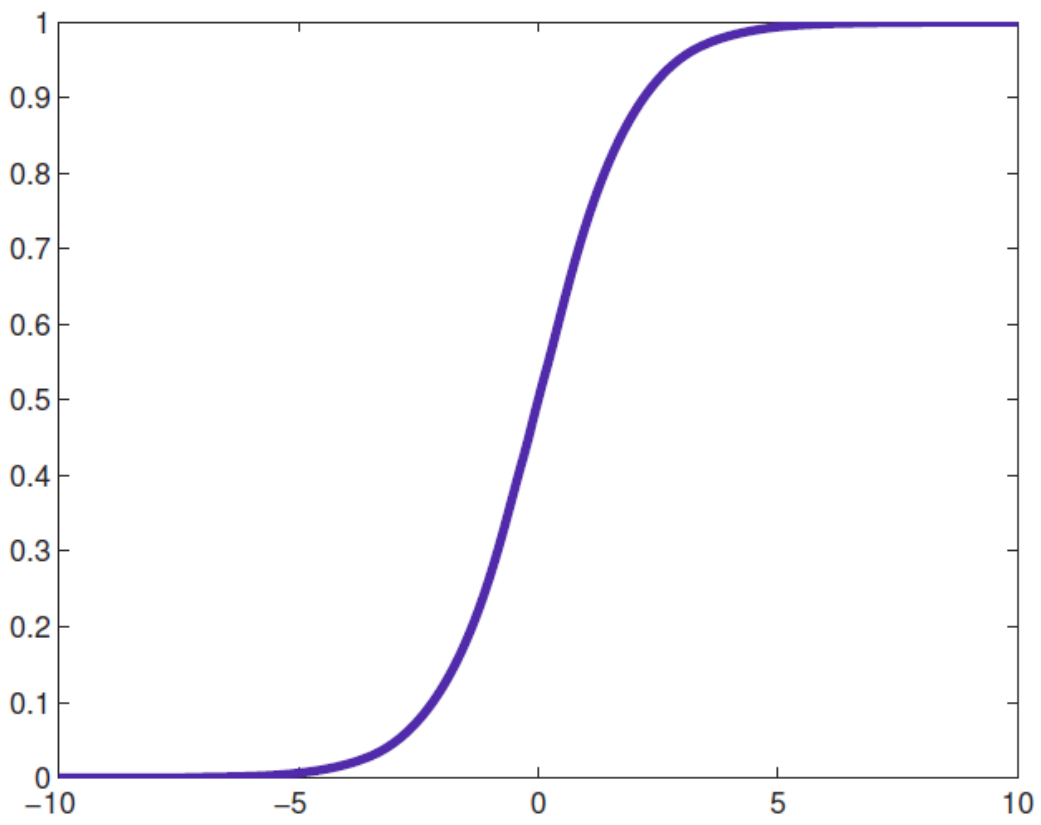

*Figure 2: Shape of a sigmoid function*

$$sigm(\theta) = \frac{e^{\theta}}{e^{\theta} + 1} \quad (3)$$

$$\min_{w} \sum_{i=1}^{N} \log\left(1 + \exp\left(-y_i \vec{w}^T \vec{\Phi}(\vec{x}_i)\right)\right) \quad (4)$$

Equation 4 does not have a simple close-form solution like linear regression, instead numerical optimization techniques like gradient descent, Newton's method, or the Levenberg-Marquardt algorithm can be utilized to learn the weights (Nocedal & Wright, 2006).

### 2.2.2 K-Dimensional Tree

K-Dimensional (K-D) Trees are one of the most popular nearest neighbour classification algorithms. In low dimensions D, (i.e. D < 20; in other words when only a handful of financial factors are used for bankruptcy prediction), a K-D Tree can be a very efficient algorithm even when training the classifier with a large number of companies (N > million). The main reason for such efficiency lies in its binary tree structure, which sub-divides the company data as being above or below a hyperplane defined by the median in all dimensions (Figure 3). Another nice property of the K-D Tree model is that it is a non-generalizing method (i.e. it simply memorizes all the previous examples without learning any parameters) and consequently can handle any probability distribution and nonlinearity in the model. Most of the computational efforts of this method lie in the construction of the tree structure. Once the search tree has been set up based on the training samples, classifying a new company as either bankrupt or not is just a matter of retrieving its k-nearest neighbours and using a majority voting scheme to decide on the label. Intuitively this makes sense: if most companies with similar financial factors are not going bankrupt, then it is most likely that this new company under investigation will also not go bankrupt, and vice versa.

The most commonly used distance metric for measuring the closeness between two companies is the Euclidean norm. Compared to a brute force approach of finding the k-nearest neighbours, a K-D Tree can reduce the computation complexity from $O(DN^2)$ to $O(log(N))$. However, this reduction is only achievable when dimension D is sufficiently low; in very high dimensional space, the efficiency of a K-D Tree diminishes due to the curse of dimensionality. Even though a K-D Tree does not learn any parameters, it still has a parameter that needs to be tuned: the number of neighbours 'k' to search. A larger 'k' can usually reduce the effect of random noise, but the boundaries become less distinct. A smaller 'k' can decrease the bias at the expense of increasing the variance. Hence, choosing 'k' is effectively balancing the bias-variance dilemma/trade-off.

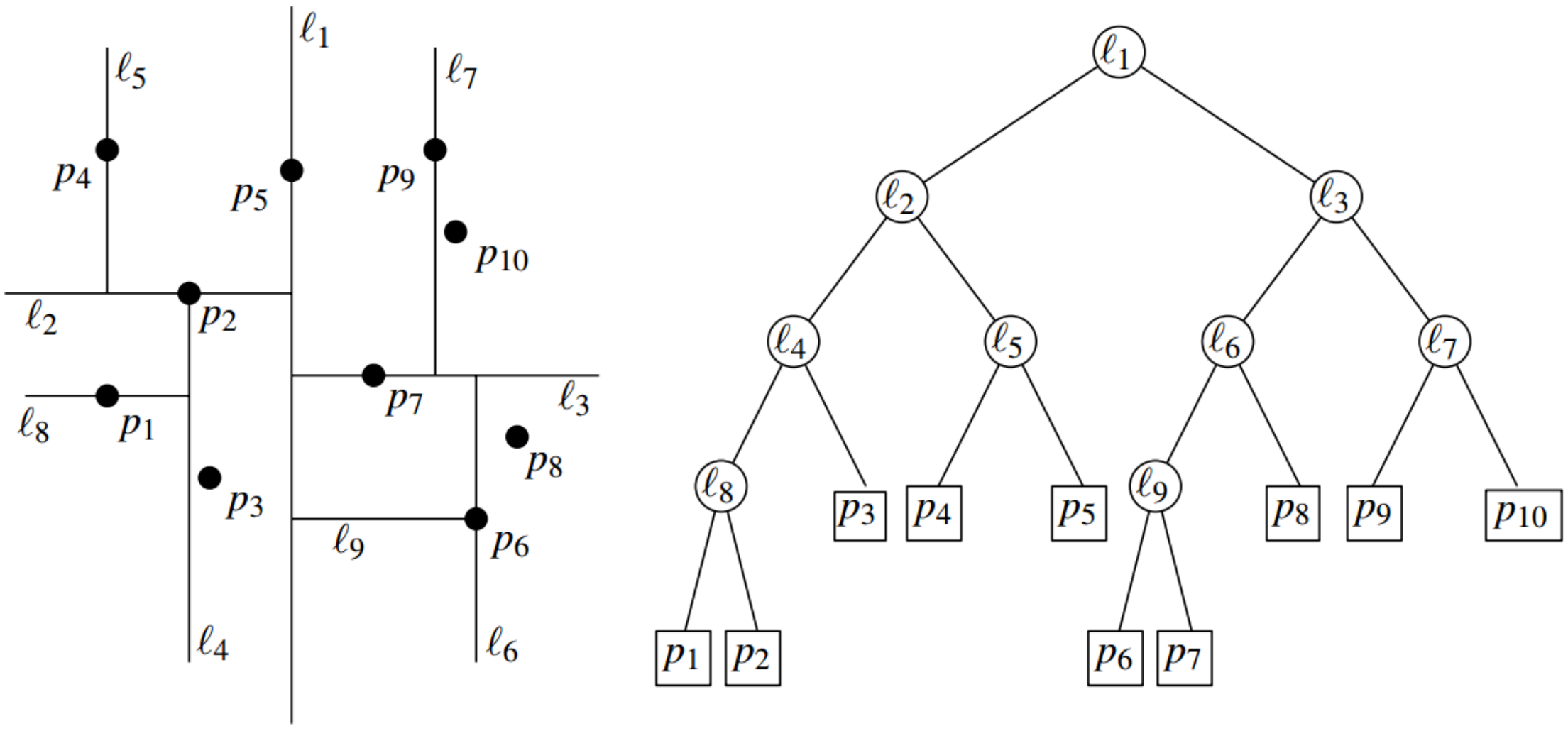

*Figure 3: Graphical representation of a 2D K-D tree. The left side shows the spatial partitioning (with hyperplanes $\ell_1, \ell_2, \dots \ell_9$ drawn) and the right side shows the corresponding tree structure.* (Berg, Cheong, Kreveld, & Overmars, 2008)

### 2.2.3 Support Vector Machine

Among the various Kernel Machines, support vector machine (SVM) is one of the most widely used for classification tasks. The mathematical formulation can be quite involved, but from a geometric point of view it can be quite simple to visualize and explain. When there are two linearly separable classes (**O** and **X**) in a two-dimensional feature space, there can be many different solutions that all yield good accuracy (or similarly, minimum errors) on the training data (Figure 4). The SVM solution matches what an intelligent human agent would naturally choose as the decision boundary, i.e. the one that is the farthest away from either cluster (shown in purple). This makes the decision made by SVM more robust against errors (both random and systematic). For example, some financial ratios such as the debt-to-equity ratio rely on subjective valuations of intangible assets by accountants, which are error-prone.

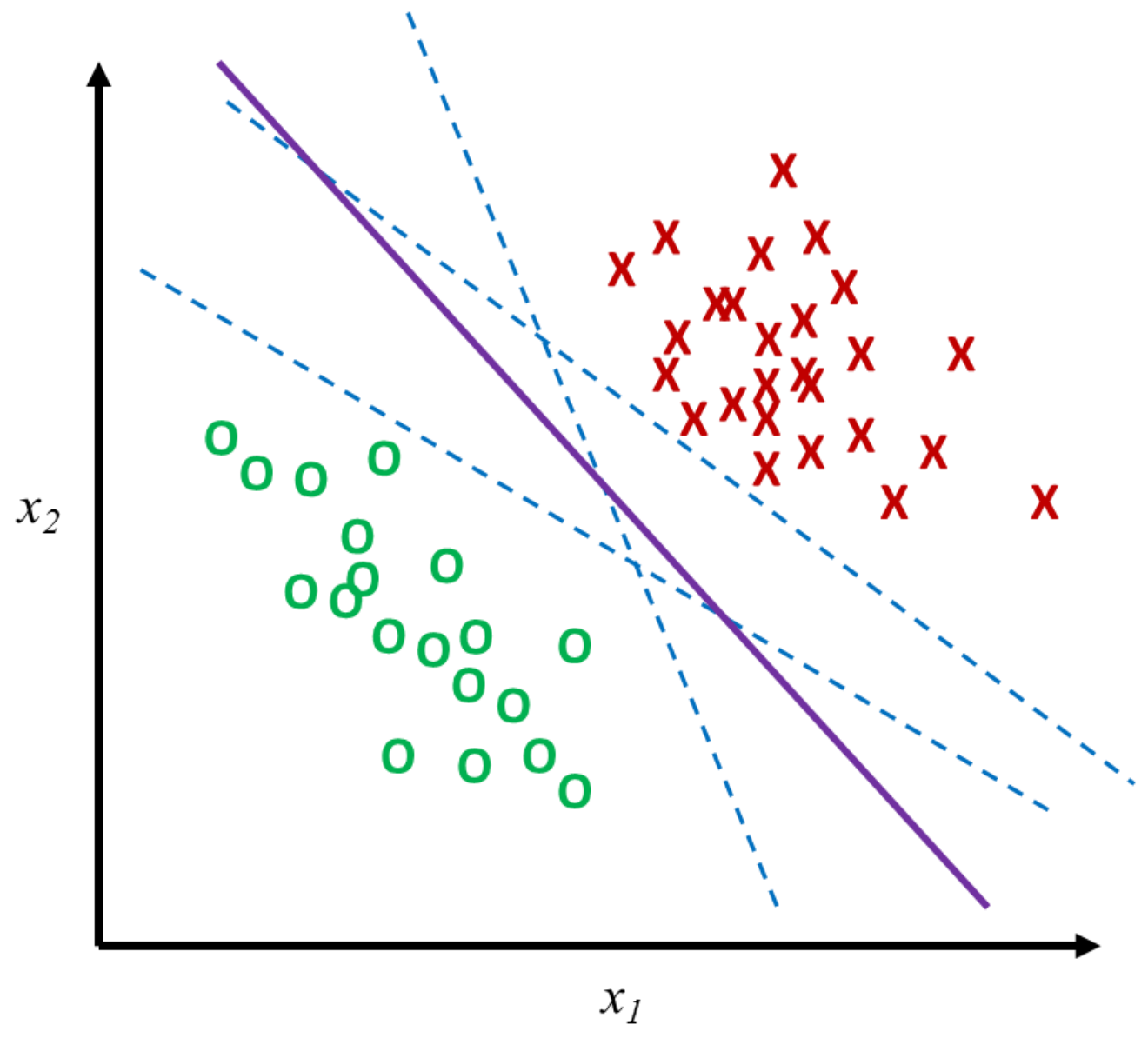

*Figure 4: Binary separation of* ***O*** *and* ***X*** *using SVM in 2D. The* ***blue*** *dashed lines indicate potential decision boundaries with zero training errors (i.e. no labels are located on the wrong side of the line) and the* ***purple*** *solid line shows the SVM decision boundary.*

This concept of choosing the decision boundary with the largest margin is one of the distinguishing characteristic of SVM. Compared to logistic regression, which is more concerned with maximizing the probability of the two classes, SVM concentrates on maximizing the separation between the support vectors and therefore the classification accuracy (i.e. minimizing the generalization error). Since only the support vectors (a subset of training points closest to the boundary) have a significant impact on the decision boundary, the solution can be considered sparse and full knowledge of the posteriori/class probabilities is not necessary; this improves the efficiency of the algorithm.

For SVM to learn from the training data, the primal problem given in Equation 5 is often converted to its equivalent dual problem, provided in Equation 6. In this case, the labels $\vec{y}$ are {-1, 1} instead of {0, 1} as in the logistic regression formulation. Although the dual problem appears to be more complicated than the primal problem, there are many advantages that reduce the complexity of this formulation. When the dimension of the input feature space (D) is much greater than the number of training points (N), the dual problem only needs to estimate N number of $\alpha$ instead of D number of $w$. Even though this is not usually the case (i.e. N is typically greater than D), only the support vectors will have a non-zero $\alpha$, therefore it can still be more efficient. Furthermore, $\Phi(x_i)^T\Phi(x_j)$ can be kernelized into $K(x_i, x_j)$ and calculated directly, which bypasses the need to explicitly solve for $\Phi(x)$. Finally, because this formulation of the objective function is convex, a globally optimal solution can always be determined (i.e. the algorithm would not converge to a local optimum). Note that C in the equations is a hyper-parameter that sets the degree of regularization. This allows some "slack" in the model in cases where there are outliers and the two classes are not perfectly separable. When C is large, a hard and narrow margin is obtained between the two classes, while a small C returns a soft and wide margin.

$$\min_{w} \ \vec{w}^T\vec{w} + C\sum_{i}^{N}\max\left(0, 1 - y_i\left(\vec{w}^T\vec{\Phi}(x_i) + b\right)\right) \qquad 5$$

$$\begin{aligned} &\max_{\alpha_i \geq 0} \sum_i \alpha_i - \frac{1}{2}\sum_{jk}\alpha_j\alpha_k y_j y_k \Phi(x_j)^T\Phi(x_k) \\ &s.t. \ \ 0 \leq \alpha_i \leq C \quad \forall i \\ &\sum_i \alpha_i y_i = 0 \end{aligned} \qquad 6$$

### 2.2.4 Decision Trees

Similar to a K-D Tree, a Decision Tree has a binary tree structure and a decision-making logic that is easy to interpret. It is a supervised machine learning algorithm which is considered non-parametric (even though the thresholds at each node are being learned as parameters) because the size of the tree can grow to adapt to the complexity of the classification problem (Alpaydin, 2010). Starting with a single node where all the training data reside, the data are split into two nodes by asking a binary “if-and-else” question. For instance, if the company’s net profit in the previous quarter is below a million dollars step to the left, else step to the right. After this split, the purity of the class labels in both nodes should have improved, meaning most of the financially healthy companies are together in one node while the companies at a high risk of insolvency are in the other node (Figure 5). However, sometimes the reduction in entropy is not immediate at the current binary decision, but a few nodes later (Bishop, 2006). These “if-and-else” questions are continuously being asked for each input feature, resulting in the growing of the tree. Once the purity at the leaf nodes satisfy some quality measures such as the Gini index and/or cross-entropy, this branching process can stop, and pruning can be performed to simplify the tree by trimming and merging less informative nodes (i.e. remove questions that did not provide significantly new insights).

When training is done and a new company has been classified by this decision tree, an analyst or manager can trace all the questions answered by this company’s features (e.g. financial ratios) to potentially pinpoint the strengths and weaknesses of this company which resulted in it being assigned to a particular group. For example, if the company is labelled as being a candidate for insolvency, traversing through the nodes of this company might indicate their gearing ratio is too high for their business. Unlike black-box machine learning techniques such as Artificial Neural Network, a decision tree can provide an easy to understand explanation for its reasoning/decision. Unfortunately, the binary decisions being localized (axis aligned) to a single feature also has its disadvantage: since the features are orthogonal to each other in feature space, a straightforward linear decision boundary at 45 degrees would require many nodes.

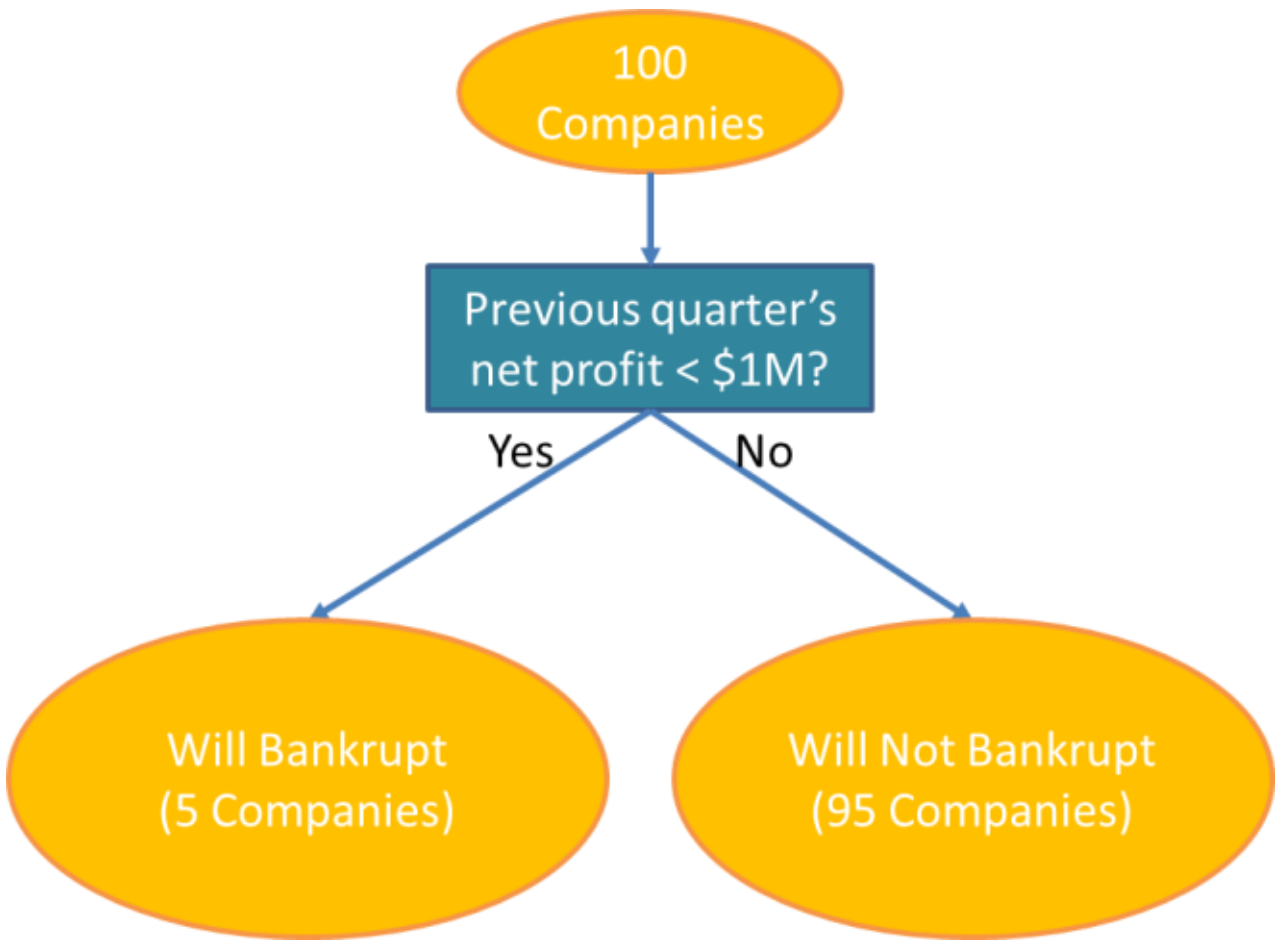

*Figure 5: Decision tree using one level for binary classification*

### 2.2.5 AdaBoost

Motivated by the idea that a committee can make a better judgement than a single individual, adaptive boosting (AdaBoost) is a technique where decisions of many weak learners (e.g. decision trees with only one or two levels) are combined to form a consensus that mimics a strong learner (Freund & Schapire, 1997). This has some resemblance to the method of Bootstrap Aggregating (a.k.a. bagging) in machine learning, where the variation in decisions from different committee members are left to chance by only submitting a subset of randomly drawn training samples with replacement to the various weak learners. The individual decisions are later combined to form the final decision. But bagging and boosting are two very different methods. In AdaBoost, the first base classifier uses a uniform weight for all training samples. All subsequent weak learners use the previous classifier's error as weights, where the weights of the improperly labelled points are increased. If these points are mislabelled again in training their weights will be increased even more for subsequent weak learner, otherwise their weights are decreased.

To put things into context, imagine a team in the human resources department deciding whether to employ an applicant or not. In the case of bagging, each employee will see slightly different information about the applicant (e.g. employee A might be able to see their volunteering experience and education background, and employee B might see their volunteering experience and work experience). The team members will independently come up with their recommendation, and the committee's final hiring decision would reflect all individual team members' decision. In the case of AdaBoost, employee A would look at the complete résumé and make recommendations. These recommendations will then be forwarded to employee B, who will make recommendations based on the full résumé and employee A's decisions. By the time the reports have circulated around the committee, the entire teams' feedback would have been incorporated into the final hiring decision.

### 2.2.6 Artificial Neural Network

As the name suggests, neural networks are inspired by the human brain. Individual neurons in our brains make simple decisions and yet together they control our complex motor functions, cognitive abilities, etc. The individual neurons can be mathematically approximated by logistic regression, and therefore an artificial neural network (ANN) can be thought of as multiple layers of connected logistic regression classifiers. Figure 6 illustrates the typical structure of a two-layer neural network known as the Multi-Layer Perceptron. An ANN with this architecture can already be considered a universal approximator, having the capability/flexibility to model any continuous functions. With modern day computers, the trend is to have more hidden units and hidden layers, giving it a deep network architecture. Although an interconnected network with many edges and nodes can be daunting at first, when broken down into its fundamental building blocks an ANN is just calculating the weighted sum of several input features. If this value is larger than a threshold, the activation function "fires" (i.e. returns a value of one instead of zero).

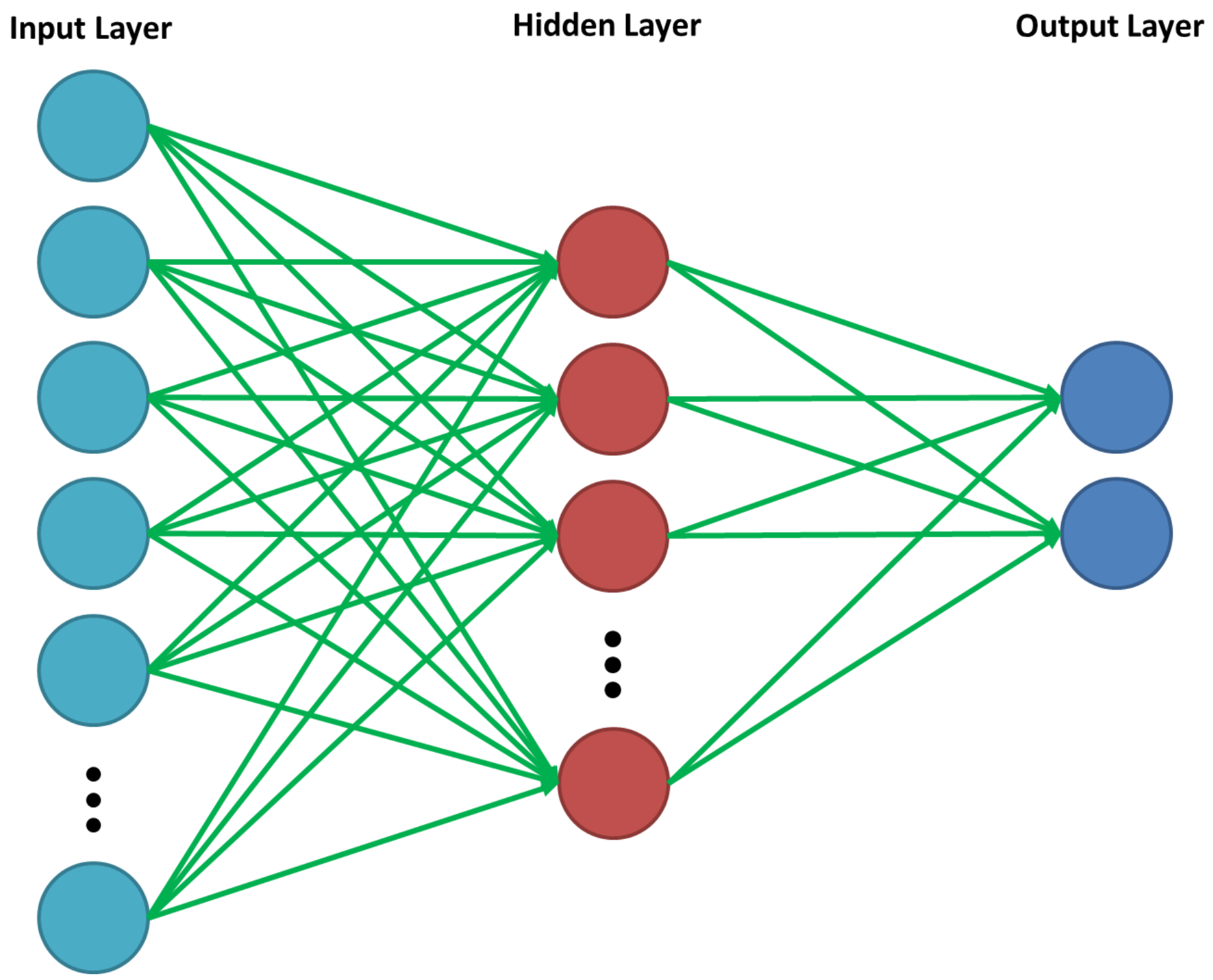

*Figure 6: Structure of a fully connected two-layer artificial neural network*

ANN share similarities with a lot of the other machine learning algorithms already discussed. For example, it combines many simple decisions into a complex nonlinear decision boundary. However, unlike SVM (which is convex), ANN is sensitive to the initial parameters and can often get trapped in a local minimum. But one noteworthy benefit of ANN is that it supports sequential learning; this means the ANN can continuously update itself as new data becomes available over time without having to re-train the entire model from scratch like SVM and K-D tree. This attribute of ANN can be critical in stock market forecasting. Also, ANN can be faster than SVM because the model is generally more compact.

### 2.2.7 Gaussian Processes

Gaussian process is an interesting machine learning method that has been gaining popularity in bankruptcy prediction in recent years. One of the main reasons why it was not widely adopted in the past is because of its high learning complexity (Equation 7), which is $O(N^3)$. But with modern day computers and improved numerical approximations, this bottleneck is slowly disappearing. Not only is Gaussian process a Bayesian non-parametric regression and classification technique (meaning it can be infinitely flexible to fit any data without over-fitting), it assigns proper probability measures to all outputs. Furthermore, it is comparable to a single layer neural network with an infinite number of neurons when used with a specific kernel (Equation 8) (Rasmussen & Williams, 2006). The ability to not over-fit is related to the principle of Occam's razor being natively built into the maximum likelihood formulation of GP. As training and model selection can be done simultaneously, the decision boundary can be as complex as the data allow but no further. Perhaps the biggest advantage of GP for bankruptcy forecasting is the ability to give a probabilistic confidence level of the prediction. For instance, if both companies A and B are being classified as candidates for insolvency with a 51% and 99% probability, respectively, then we can further say company A is in a better shape than company B and is only a borderline bankruptcy candidate, whereas company B has a very high likelihood of declaring bankruptcy.

$$\min_{\theta} \frac{1}{2}\vec{w}^T K(\theta,\vec{x})^{-1}\vec{w} + \sum_{i=1}^{N}\log\left(sigm\left(y_i w_i\right)\right) + \frac{1}{2}\log\left| I + \nabla\nabla\log\left(sigm\left(y_i w_i\right)\right)^{1/2} K(\theta,\vec{x})\nabla\nabla\log\left(sigm\left(y_i w_i\right)\right)^{1/2}\right| \quad (7)$$

$$K(x,x') = \frac{2}{\pi}\sin^{-1}\left(\frac{2\vec{x}^T\Sigma\vec{x}'}{\sqrt{\left(1+2\vec{x}^T\Sigma\vec{x}\right)\left(1+2\vec{x}'^T\Sigma\vec{x}'\right)}}\right) \quad (8)$$

# 3 OBJECTIVES

The aim of this thesis is to diagnose the financial health of businesses using machine learning algorithms. A detailed study of using several data-driven models to forecast corporate bankruptcy (firm default) was conducted. Both qualitative assessments from financial experts and quantitative econometric factors will be considered for training models for predicting financial credit risk.

# 4 METHODOLOGY

To ease comparison of the applied machine learning method to other techniques, as well as to allow validation by other researchers, secondary data was used in this project. Two publicly available datasets from the Machine Learning Repository hosted by the Center for Machine Learning and Intelligent Systems at the University of California, Irvine, were processed.

The first dataset (denoted as Dataset 1) contains 250 records of six qualitative measures of different companies from loan experts with about nine years of experience. Note in the original paper by Kim & Han (2003), they used a genetic algorithm to forecast bankruptcy of 772 manufacturing and service companies in Korea and only 250 of those records have been shared. The following six qualitative measures were established by one of the largest Korean banks (Kim & Han, 2003):

- Industrial Risk
- Management Risk
- Financial Flexibility
- Credibility
- Competitiveness
- Operating Risk

More details about these measures are provided in Appendix 1. Basically, industrial risk measures the health and future potential of the industry, management risk measures the organizational structure and managers' capabilities, financial flexibility is a measure of the company's cashflow, credibility is a measure of the company reputation and credit scores, competitiveness is a measure of the company's market position and competitive advantages, and operating risk is a measure of its efficiency in production. As presented in the original paper, the authors' proposed genetic algorithm method resulted in a classification accuracy of 94%. They also compared it to two other data-mining techniques, namely induction learning and neural networks, and reported an accuracy of 89.7% and 90.3%, respectively.

In contrast, the second dataset (denoted as Dataset 2) contains 5910 instances of 64 quantitative attributes from Polish manufacturing companies between the years 2000-2012, with some still operating companies being evaluated between 2007-2013 (Zięba, Tomczakb, & Tomczaka, 2016). 5500 of those companies did not declare bankruptcy, while the remaining 410 filed for insolvency after one year. Most of the quantitative attributes are financial ratios and econometric indicators as found in majority of existing literature. A complete list of those attributes can be found in Appendix 2.

The methods for analysing these two datasets are similar and will be explained below. Note that the difference in the quality of results between Dataset 1 and Dataset 2 can be attributed to factors such as different geographic location, different dataset size, different features, and different quality of data.

## 4.1 Pre-Processing

The range of possible values for various input features can vary drastically. For example, gross margin defined by Equation 9 will always be less than one (i.e. below 100%) due to normalization, whereas some financial measures like working capital can theoretically take on any real value (i.e. negative infinity to positive infinity). To complicate the situation further, some scholars have proposed including additional features such as corporate governance structures and management practices in the prediction model (Aziz & Dar, 2006), which can have any arbitrary scale. Having huge variations in scales across dimensions leads to several issues in machine learning, such as a higher chance for numerical instability and saturation (i.e. a situation where some features dominate and mask the importance of some other feature(s) due to sheer magnitude).

$$\boldsymbol{Gross\ Margin} = \frac{\boldsymbol{Total\ Revenue} - \boldsymbol{Cost\ of\ Goods\ Sold}}{\boldsymbol{Total\ Revenue}} \qquad 9$$

One possible solution is to standardize each feature such that all have zero mean and unit variance. To achieve this, the average can simply be subtracted from every training sample and then divided by its standard deviation (Equation 10); however, if the variance of a particular feature is very small (i.e. close to zero), then this division may have numerical issues. An alternative is to simply scale the data to be between a minimum ($\min_{desired}$) and maximum ($\max_{desired}$) value of choice: for instance, zero and one (Equation 11).

$$z_i = \frac{x_i - \bar{x}}{\sigma_x} \qquad 10$$

$$z_i = \left[\frac{x_i - \min(\vec{x})}{\max(\vec{x}) - \min(\vec{x})}\right]\left[\max_{desired} - \min_{desired}\right] + \min_{desired} \qquad 11$$

## 4.2 Dimensionality Reduction

Classification with only one, two, or even three dimensional features is generally intuitive, because the classification boundary and training data can be visualized. Unfortunately, financial distress information resides in a higher dimensional feature space. In other words, only analysing three financial ratios is insufficient to provide a clear separation between successful companies and companies that will likely experience bankruptcy in the future. In high dimensional feature space, such as the 64-dimensional Polish bankruptcy dataset, it becomes difficult for humans to “see” what is happening. However, many mathematical tools exist to reduce the dimensionality of the data; not only can this potentially provide a means to perceive high dimensional data, it also reduces the complexity of the problem, and can reduce some of the noise in the data.

One popular method for linear dimensionality reduction is Principal Component Analysis (PCA). It performs a transformation/projection that maximizes the variance along each orthogonal axis (equivalent to minimizing the loss of information) by reducing the correlation between features (Hotelling, 1933). This makes sense because even though the dimensions of many machine learning problems are high, the interesting characteristics typically lie in a lower dimensional manifold. For example, researchers have proposed adding macroeconomic measures to the existing financial ratios for financial distress prediction. As a company's operation is inevitably affected by the macro environment, part of its effect is already reflected in the company's financial performance; therefore, when macroeconomic features are included, the information they provide is not totally independent from the other features. Another example is the similarities of some financial ratios. Upon close examination of the Polish bankruptcy features, one would identify that some ratios like feature 4 (i.e. current assets divided by current liabilities) and feature 55 (i.e. current assets minus current liabilities) are closely related. A simple dimensionality reduction scheme is to manually eliminate some features that are less descriptive, but this can become tedious, subjective, and gives rise to an "all or nothing" situation (sometimes correlated features can still improve the discriminative ability of the classifier). PCA combines features in such a way that a lower dimensional representation still retains most of the information. One noticeable drawback of PCA is that a clear financial interpretation of the feature space might be lost; each new feature after PCA is a linear projection of many features (e.g. financial ratios).

Mathematically the PCA solution can be obtained in different ways, one of which is by Eigen decomposition. The magnitude of the eigenvalue can be used as an indication of the information content. If the eigenvalues are sorted in descending order then typically most information is captured in the direction of the first few eigenvectors. The last few eigenvalues would be close to zero and the elimination of these last few projected features would result in only a small loss of information.

PCA is an unsupervised method for dimensionality reduction. However, if the training data contains labels, it can be advantageous to include that information when projecting the data into a lower dimensional subspace. After all, besides reducing the complexity of the following machine learning algorithms by working in a lower dimensional subspace, one of the objectives of this projection should be to maximize the separation between the different classes. One of the methods to achieve this is the Linear Discriminate Analysis (LDA). Class labels are required, thus LDA can be considered a supervised version of PCA.

Both PCA and LDA perform a linear projection. When the projection is nonlinear we can estimate the geodesic distance along the manifold and apply multi-dimensional scaling. This method is known as Isometric Feature Mapping (ISOMAP). ISOMAP is an unsupervised dimensionality reduction method that can handle the nonlinearities at the expense of more computation efforts. An alternative to this approach is to apply the "kernel trick" adopted in SVM to PCA; this yields a method known as Kernel PCA. If the kernel is chosen to be nonlinear (e.g. a radial basis kernel), then the projection will be nonlinear.

## 4.3 Learning from Data and Model Selection

Different machine learning algorithms have different methods of constructing a model of the real world using the provided data. For instance, a K-D tree learns from the data by partitioning the data and forming a binary tree structure for fast query, while a logistic regression learns by estimating some weight parameters in an optimization framework where some likelihood function is being maximized. In general, this learning from data can either be parametric or non-parametric. Parametric methods will learn some unknown parameters of a model and forget about the data (e.g. logistic regression), while non-parametric methods like GP will have to store all the training data. But even if the method is non-parametric, there will still be some optional tuning for optimal performance. Most machine learning methods presented have some hyper-parameters which are used to change the behaviour of the models. For example, in a GP classifier with a radial basis function kernel, the length needs to be set. This parameter controls the neighbourhood size when forming the decision boundary. In general, there are three ways of selecting the “best” model: (1) manual tuning by an expert, (2) cross-validation (CV), and (3) Bayesian statistics.

In the first case, a machine learning or econometrics expert would change a hyper-parameter, re-train the model, and analyse the results until a satisfactory solution is achieved. This can be time-consuming and subjective.

An automatic method is to do cross-validation either using a validation dataset or the k-fold scheme. Unless there is an abundance of datasets, doing k-fold cross-validation is typically a better choice because the same data used for training can be used for model selection. In k-fold cross-validation the training dataset is randomly clustered into ‘k’ groups. Group one is used as the validation dataset while all the remaining data is used for training. This process is then repeated with group 2 acting as the validation dataset and all other data is used for training. K-fold cross-validation terminates when all ‘k’ groups had a chance to play the role of a validation dataset. At this point, the model with the highest score or the average of all k-folds can be used to select the best model. For example, this strategy can be used to tune the softness of the margin in SVM, i.e. the ‘C’ hyper-parameter can be set using cross-validation. Figure 7 shows the cross-validation score for various choices of ‘C’. For this dataset, a hard margin appears to deliver a higher mean score, with the highest CV score around a C value of 12.

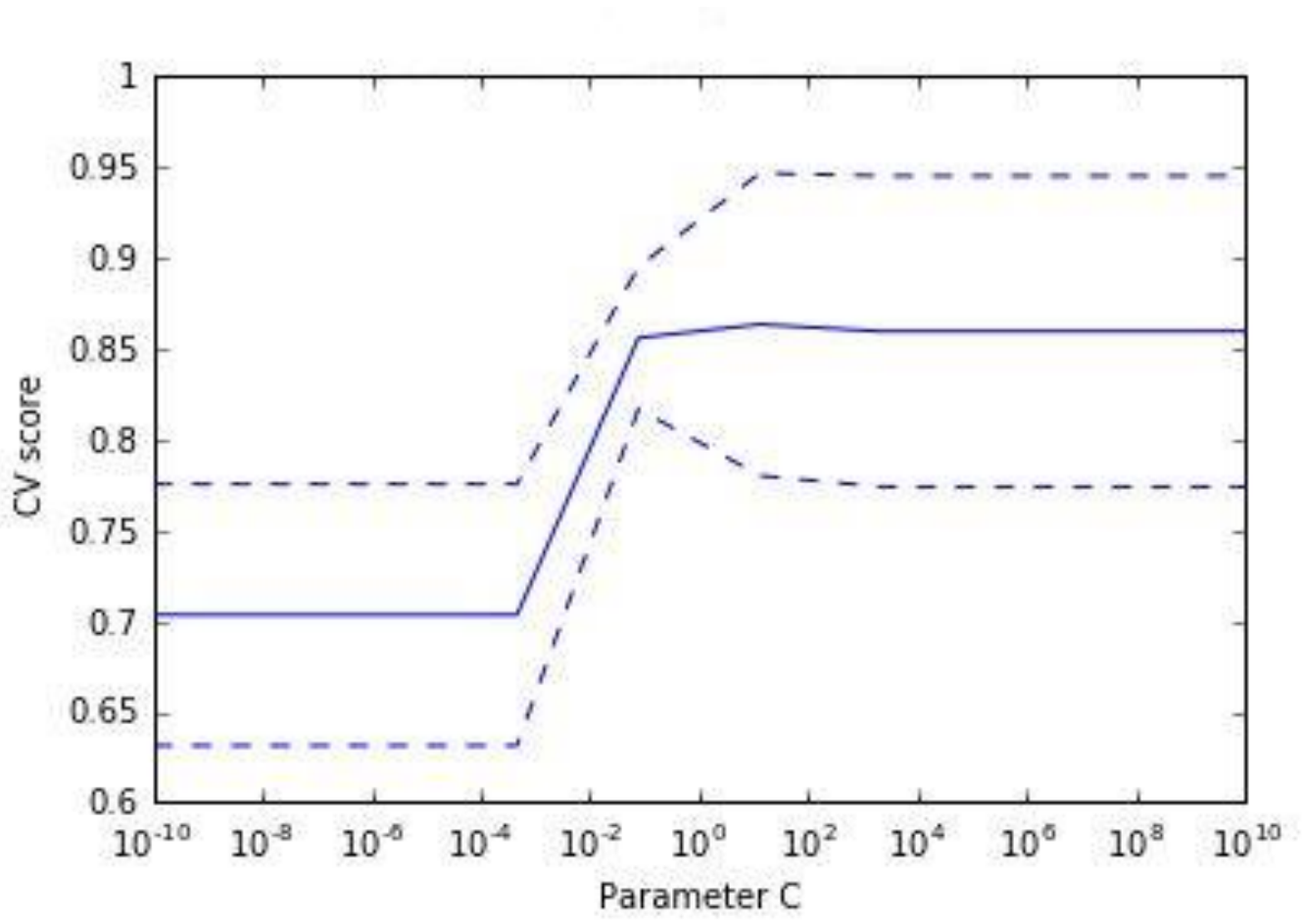

*Figure 7: Tuning of the regularization parameter of a SVM using cross-validation scores. The solid-line shows the mean of a 10-fold cross-validation and the dashed lines show the standard deviation.*

If the machine learning model is probabilistic like Gaussian Mixture Models, then a prior distribution can be assigned to find the optimal hyper-parameters. This is similar to a regularizer that sets the optimal model complexity to prevent over-fitting. One of the benefits of this approach is that model selection and training can happen simultaneously. At every iteration during numeric optimization, the hyper-parameters are updated to improve the posteriori distribution. However, not all machine learning methods are probabilistic, such as SVM and K-D tree. Therefore, to have a unified framework for model tuning, cross-validation will be used when comparing different models.

## 4.4 Accuracy Assessment

The machine learning models described in Section 2.2 can have very different characteristics and behaviour, making it difficult to judge which model is performing better. Therefore, a consistent set of tools applicable for assessing the performance of any machine learning model is important. Some of the most popular quality control measures in machine learning are defined below.

The most popular quality assessment method is the accuracy score: given a sample with ground truth classes/labels, the predicted labels from a trained model are compared to the reference labels (Equation 12). Care must be taken to ensure this testing dataset was never exposed in any of the pre-processing or training stages in the machine learning pipeline. Also, the testing set should have the same probability distribution as the training set. This can be achieved by randomly choosing a subset of points from the original dataset (e.g. 70% use for training, and 30% used for testing). This single scalar value indicates how well a machine learning algorithm can label companies in a non-recoverable financial crisis as bankrupt and financially healthy companies as not bankrupt.

$$\boldsymbol{accuracy} = \frac{\mathbf{1}}{\boldsymbol{N}} \sum_{\boldsymbol{i}}^{\boldsymbol{N}} \boldsymbol{\delta}(\boldsymbol{y}_{\boldsymbol{i}}^{\boldsymbol{pred}} - \boldsymbol{y}_{\boldsymbol{i}}^{\boldsymbol{true}}) \qquad 12$$

where $\boldsymbol{\delta}(\boldsymbol{x} = \mathbf{0}) = \mathbf{1}$ and $\boldsymbol{\delta}(\boldsymbol{x} \neq \mathbf{0}) = \mathbf{0}$

Besides the accuracy score, another set of metrics commonly used for quality control in machine learning is the precision, recall, and F1 score. Precision is a measure of how well the algorithm can find true positives (Equation 13). In the case of this thesis, it can be translated into how well the model can predict a company as bankrupt when it is actually going bankrupt. For example, a precision of 100% means that corporations flagged as bankrupt will surely experience bankruptcy in the future with great certainty. Another closely related concept to precision is recall, defined in Equation 14. Recall is a measure of how reliable can the classifier identify all true positive samples. For instance, a recall of 50% would indicate that half the bankruptcy candidates have been found while the other half of bankruptcy facing firms were missed by the classifier. Ideally, a good classifier should maximize both precision and recall, unfortunately in reality precision and recall is often a trade-off an econometrics expert would have to make while training the model. As shown in Figure 8, as the recall increases (x-axis), the precision decreases (y-axis), and vice versa. If the objective of the prediction is to highlight all companies susceptible to bankruptcy for further screening by a financial officer then a high recall would be desirable at the expense of a lower precision, because the financial officer can then manually remove the false positives. However, if the objective is to automatically reject all loan applications from companies that are near bankruptcy to avoid wasting the bank's resources to perform a more thorough interview and assessment, then a high precision (lower recall rate) might be more suitable. F1 score is defined as the weighted harmonic mean of precision and recall. It is a single number that further aids the decision making process of model comparison. A perfect model would have a F1 score of 100%, which corresponds to 100% precision and 100% recall. Thus, in general a high F1 score is preferred.

$$\boldsymbol{precision} = \frac{\boldsymbol{true\ positives}}{\boldsymbol{true\ positives} + \boldsymbol{false\ positives}} \tag{13}$$

$$\boldsymbol{recall} = \frac{\boldsymbol{true\ positives}}{\boldsymbol{true\ positives} + \boldsymbol{false\ negatives}} \tag{14}$$

$$\mathbf{F1}\ \boldsymbol{score} = \frac{\mathbf{2}\ \times \boldsymbol{precision}\ \times \boldsymbol{recall}}{\boldsymbol{precision} + \boldsymbol{recall}} \tag{15}$$

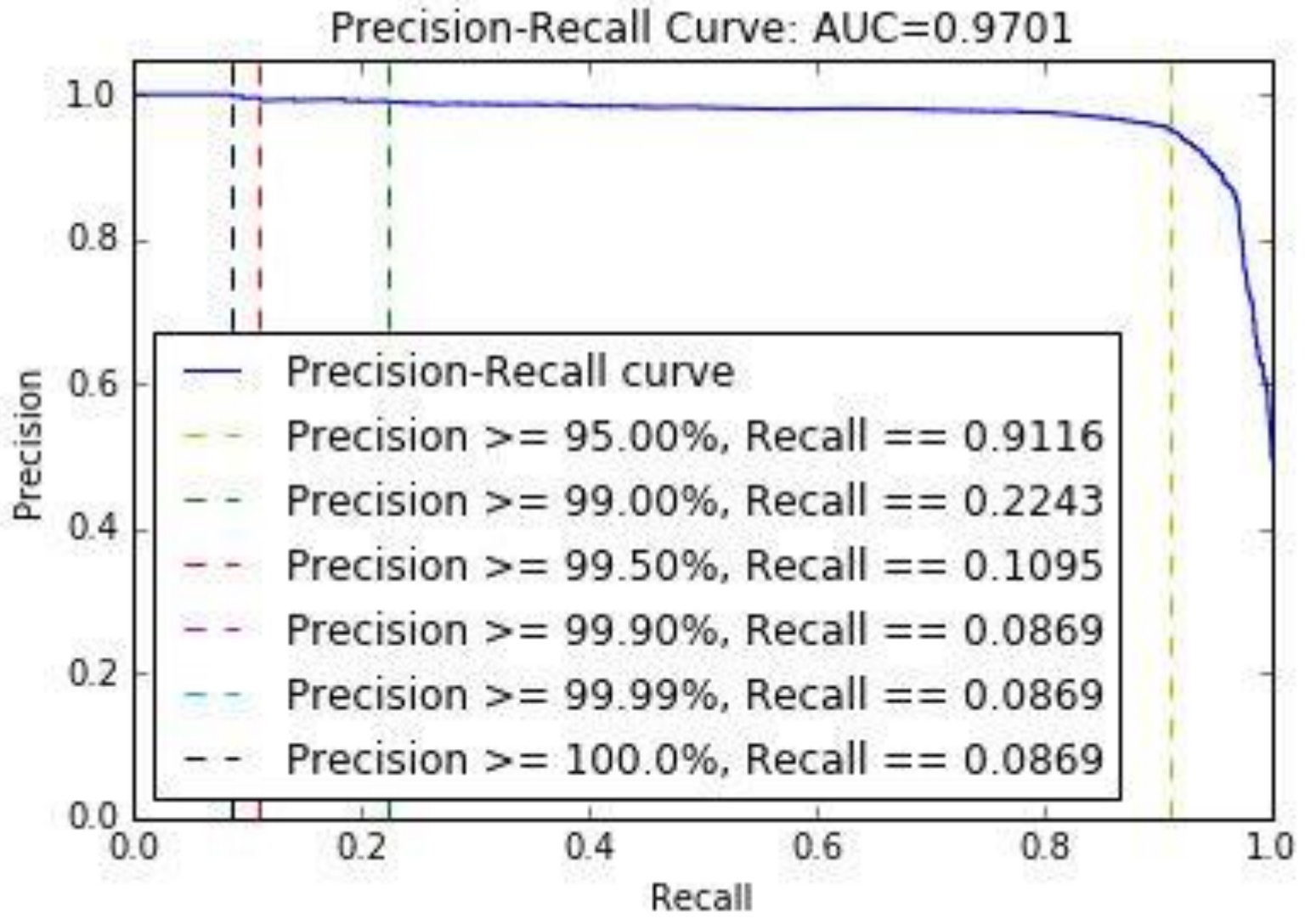

*Figure 8: Inverse relation between precision and recall*

Another way to visualize and select the trade-off between precision and recall is to analyse the Receiver Operating Characteristic (ROC) curve (Figure 9). Depending on the threshold value chosen given a boundary curve, the true positive rate and false positive rate can be manipulated. However, because they are correlated, a low threshold value will not only ensure a high positive rate, it will also give a high false positive rate. Therefore the optimal threshold value is typically in the top left-hand corner of the curve where the true positive rate is much higher than the false positive rate. To compare different models using this approach, the area under the ROC curve can be calculated. The closer the area is to one, the better the discriminant power of the classifier.

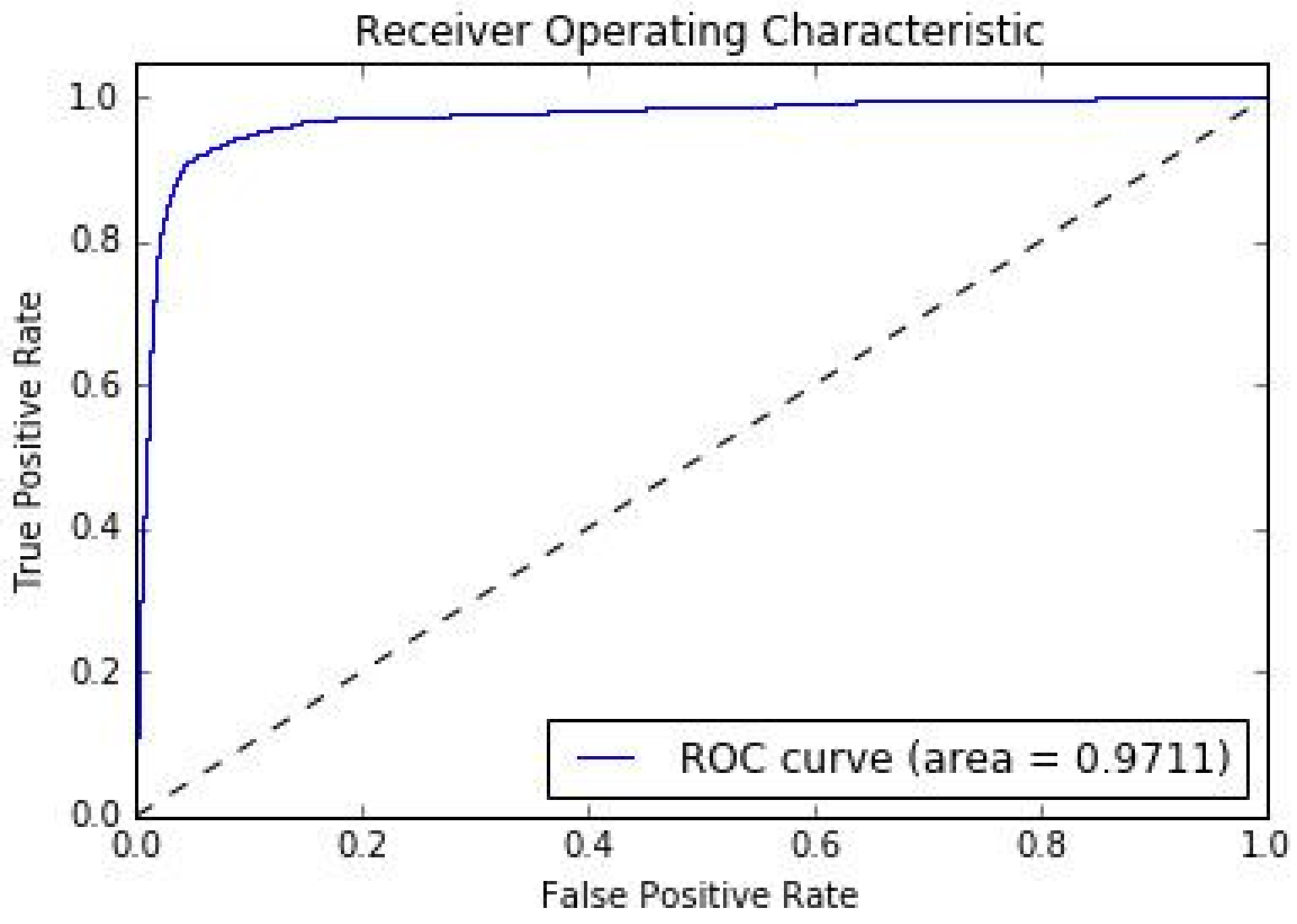

*Figure 9: ROC curve of a classifier shown in solid-line and random guesses shown in dash-line*

The relationship between true positives, false positives, false negatives, and true negatives can be summarized more thoroughly in a simple confusion matrix as well (Figure 10). This gives the probability of the four scenarios in a binary classification problem in an easy to visualize manner. In this figure, the true positive rate is 96.6%, false positive rate is 3.4%, false negative rate is 4.8%, and the true negative rate is 95.2%.

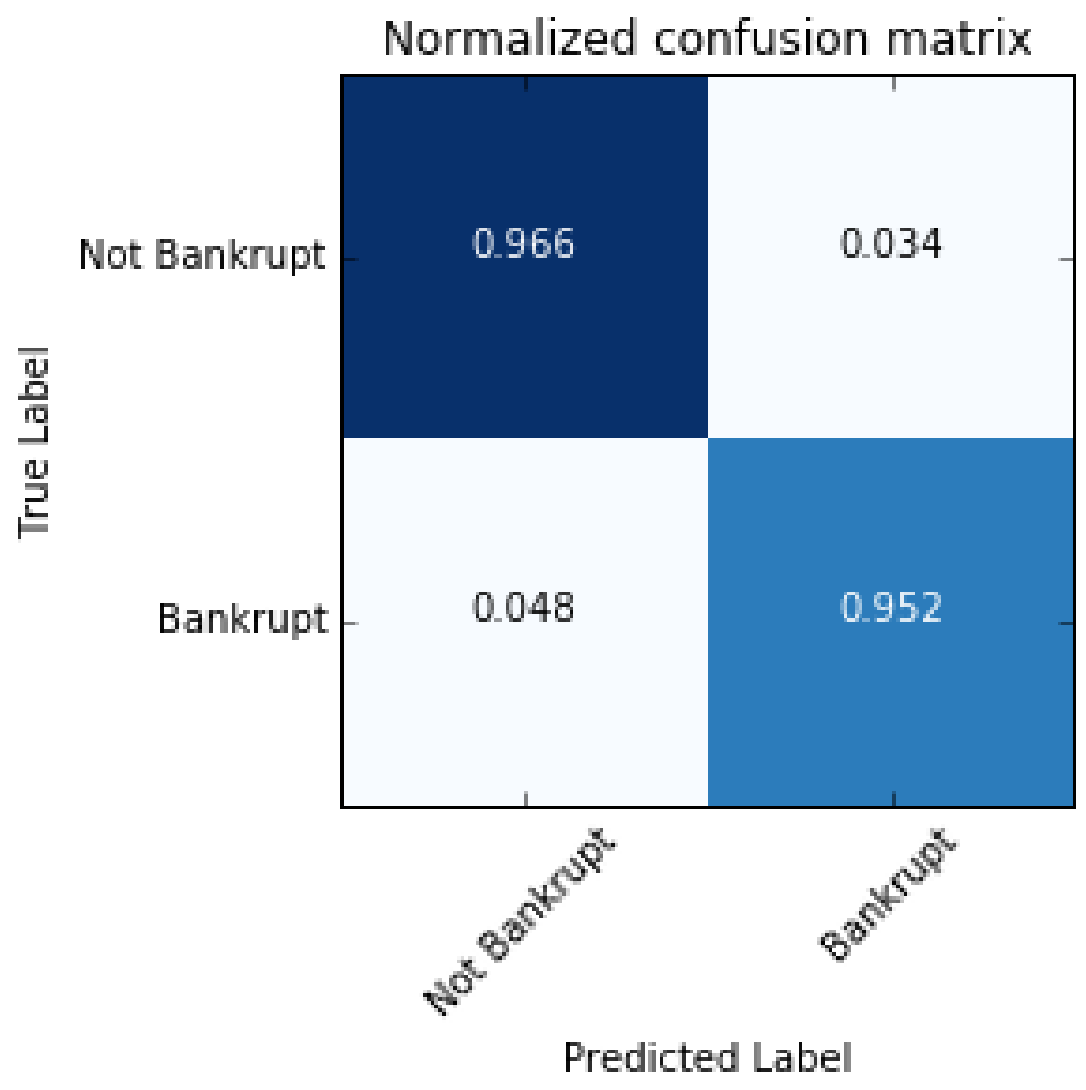

*Figure 10: Confusion matrix of a binary classifier normalized between zero and one*

# 5 Results and Analysis

For both Dataset 1 and Dataset 2, appropriate pre-processing steps have been taken to ensure all the features are scaled between negative one and positive one. In order to have ground truth data for assessing the accuracy of various classification models the datasets have been split into 80% training and 20% testing. The 20% of testing data is never used in any of the machine learning steps except in the final accuracy assessment stage. In using this method, the possibility of over-fitting and introducing personal bias when choosing model parameters can be reduced.

## 5.1 Dataset 1: Korean Corporate Bankruptcy

### 5.1.1 Visualization of the Data

Human beings' visual cortex are incapable of understanding six-dimensional space, therefore to be able to recognize patterns in the data visually the features need to be projected onto a lower-dimensional sub-space; the most intuitive sub-space being two-dimensional (Figure 11) and three-dimensional (Figure 12). Note, the data that has been labelled bankrupt is denoted in orange and non-bankrupt data is denoted in purple. It can be perceived that all four dimensionality reduction techniques are able to separate the two clusters well. In particular, the post-projection data using PCA, LDA, and Kernel PCA are linearly separable. Although a clear boundary between bankruptcy and non-bankruptcy samples can be perceived from the ISOMAP results, the separation boundary is nonlinear. Based on visual assessment, a significant benefit of using three features instead of two features cannot be perceived. Therefore, the two-dimensional sub-space will be chosen for simplicity and better visualization of the classification results. To better understand the impact of the different dimensionality reduction methods, various classifiers will be applied to the four two-dimensional datasets.

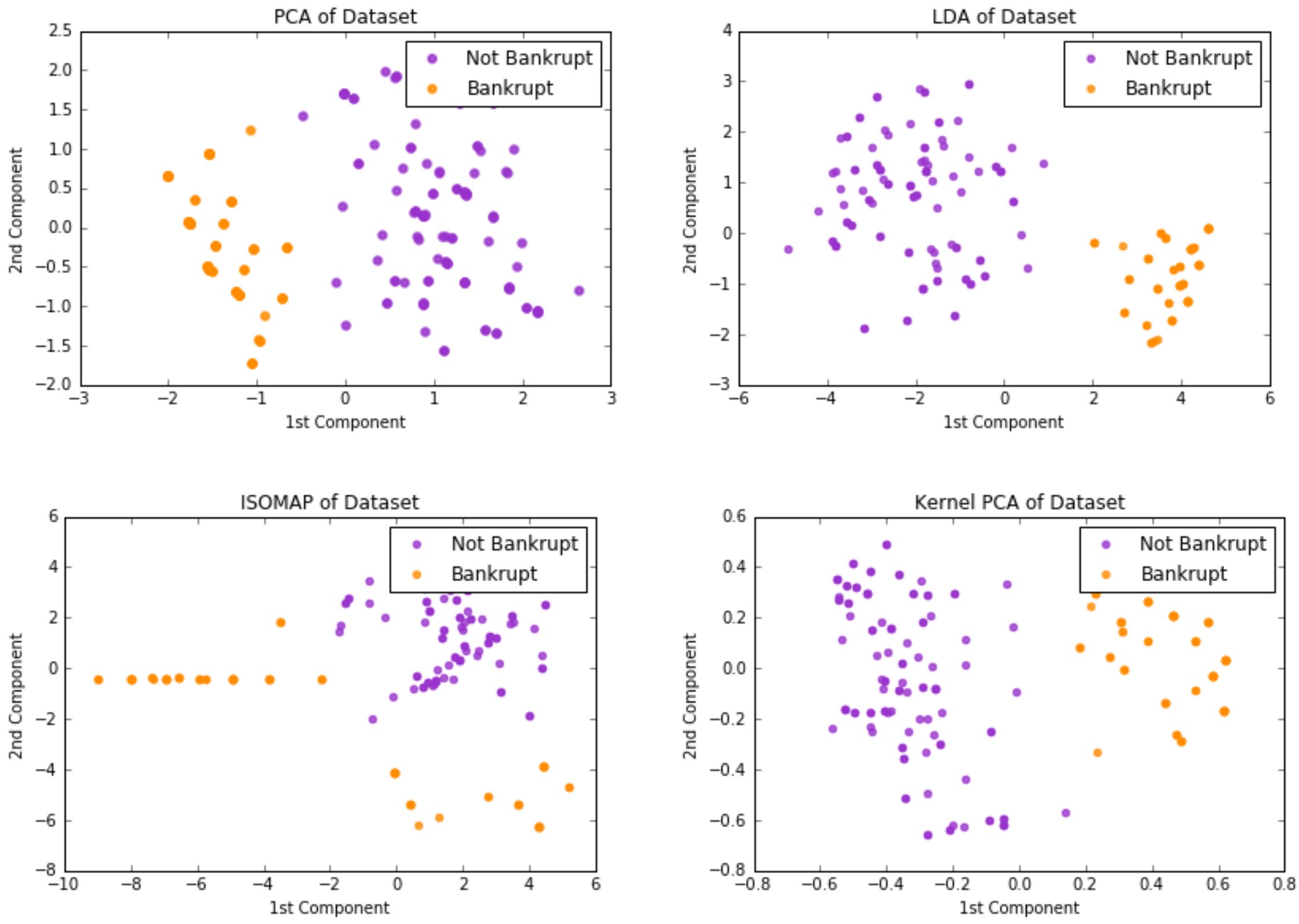

*Figure 11: Two-dimensional visualization of the Korean bankrupty dataset*

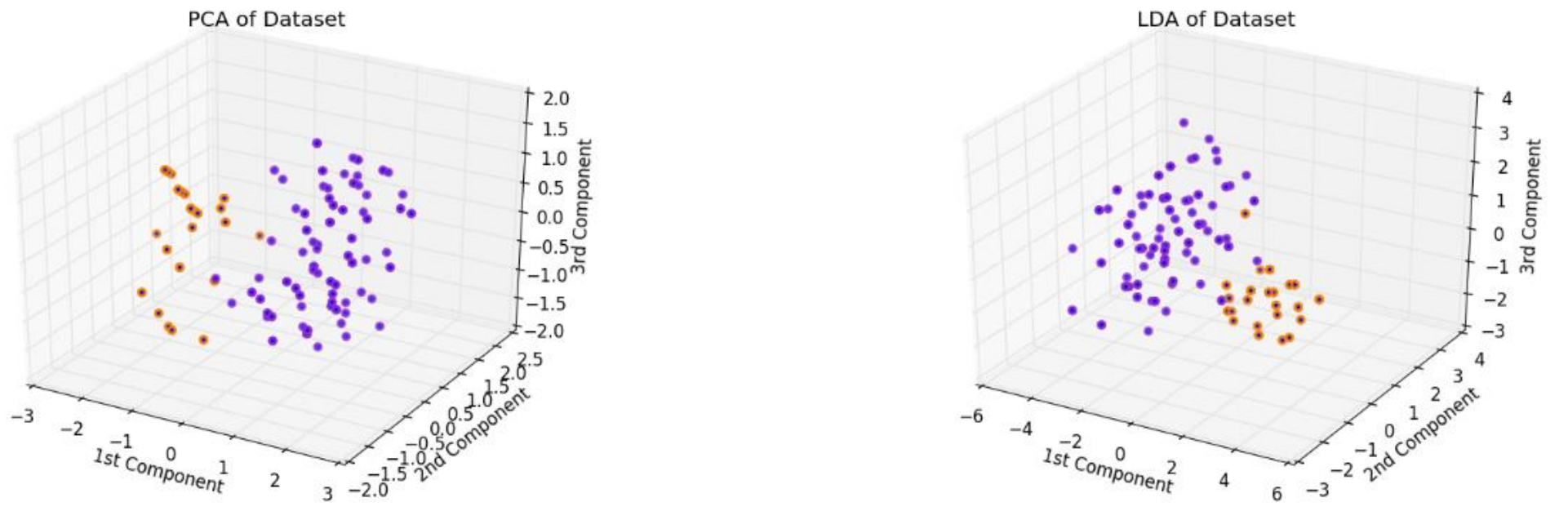

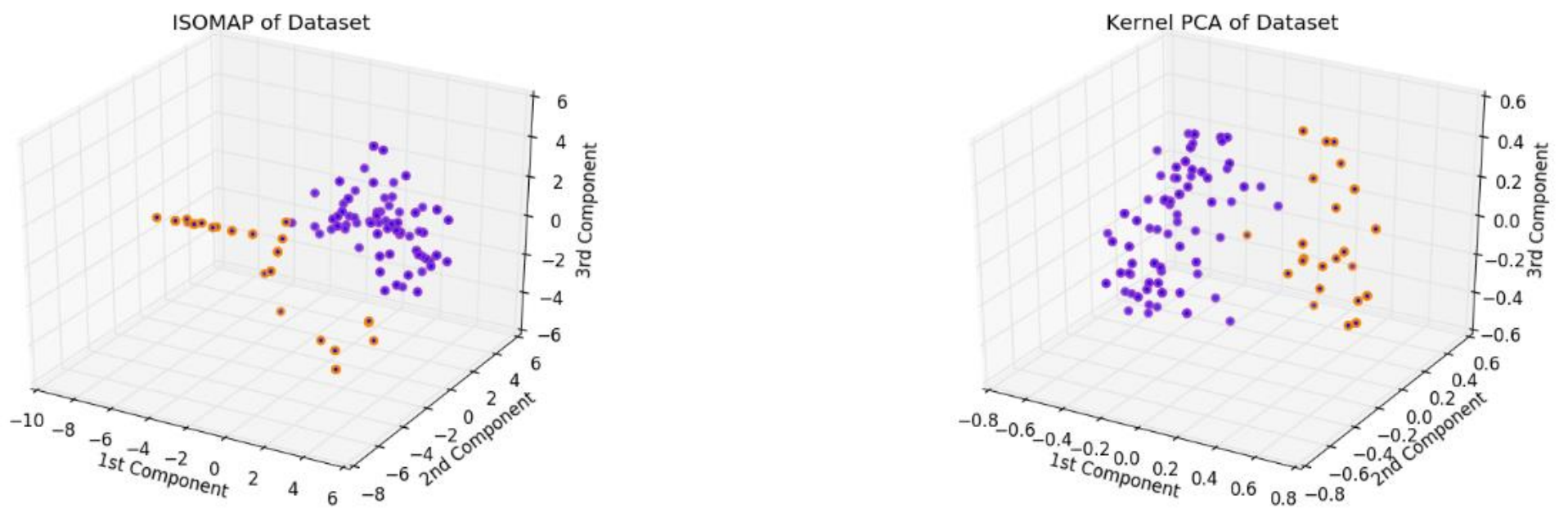

*Figure 12: Three-dimensional visualization of the Korean bankrupty dataset*

### 5.1.2 Binary Classification

Figures 13, 14, 15, and 16 illustrates the decision boundaries computed using different machine learning models on the PCA, LDA, ISOMAP, and Kernel PCA projected training and testing datasets, respectively. In the last column of these figures, their corresponding confusion matrix is also provided to showcase its classification accuracy as well as Type-I and Type-II errors. In the linearly separable sub-spaces (i.e. Figures 13, 14, and 16), logistic regression, decision tree, and AdaBoost are all able to learn a simple linear boundary. In this case, AdaBoost which is built from decision stumps (as described in Section 2.2.5) gives the same result as the decision tree classifier. The irregular decision boundaries from K-D tree, and curved boundaries from SVM, ANN, GP delivers similar level of classification accuracy, but it can be argued that it is more complex than necessary in this case.

In the ISOMAP projection scenario (Figure 15), where the data are not linearly separable in a two-dimensional space, the nonlinear classifiers outperformed logistic regression. Decision tree and AdaBoost in this case is capable of automatically learning the fact that a linear decision boundary is insufficient and separates the two-classes using different nonlinear boundaries.

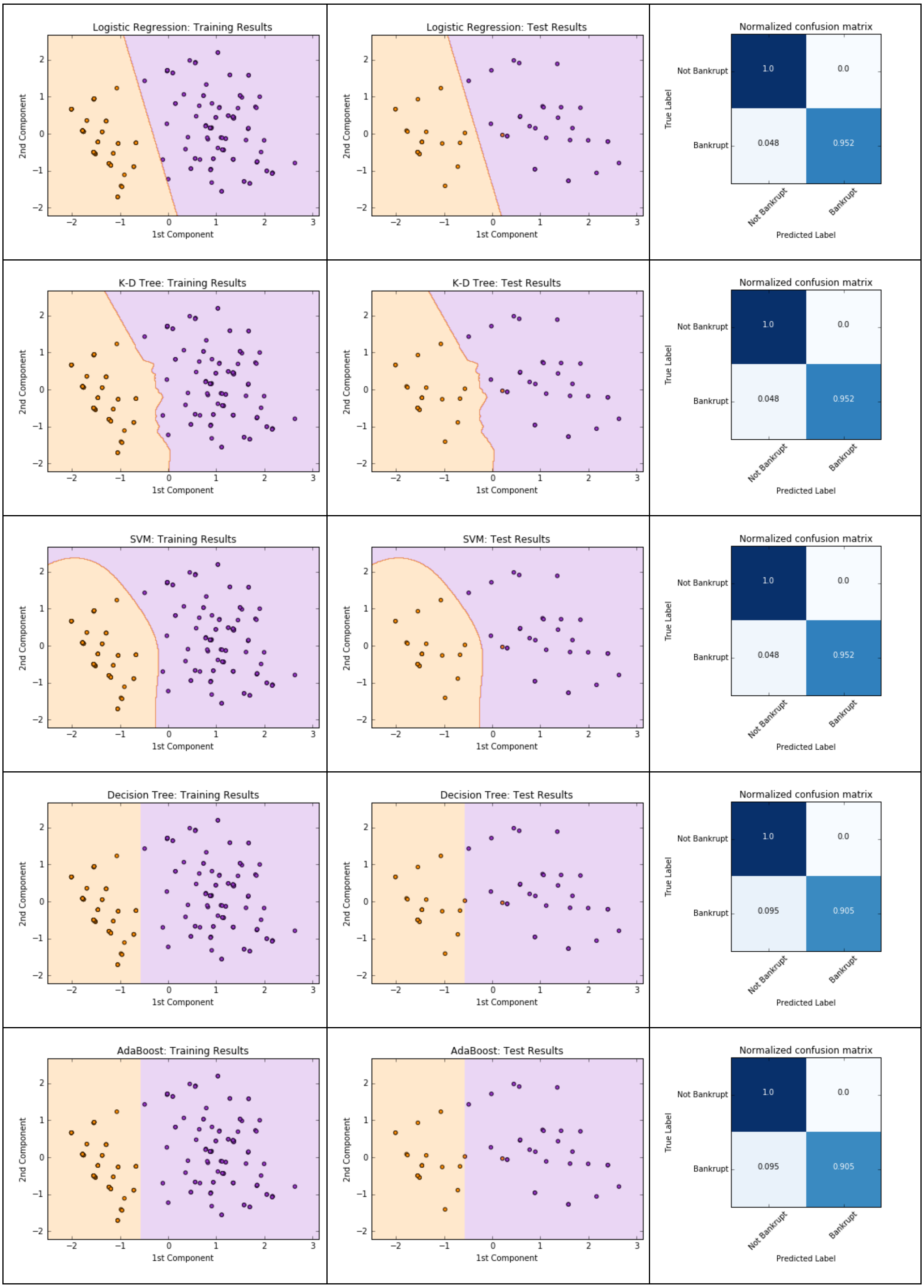
Logistic Regression: Training Results
Logistic Regression: Test Results
K-D Tree: Training Results
K-D Tree: Test Results
SVM: Training Results
SVM: Test Results
Decision Tree: Training Results
Decision Tree: Test Results
AdaBoost: Training Results
AdaBoost: Test Results
1st Component
2nd Component
Normalized confusion matrix
True Label
Predicted Label
Not Bankrupt
Bankrupt
1.0
0.0
0.048
0.952
0.095
0.905

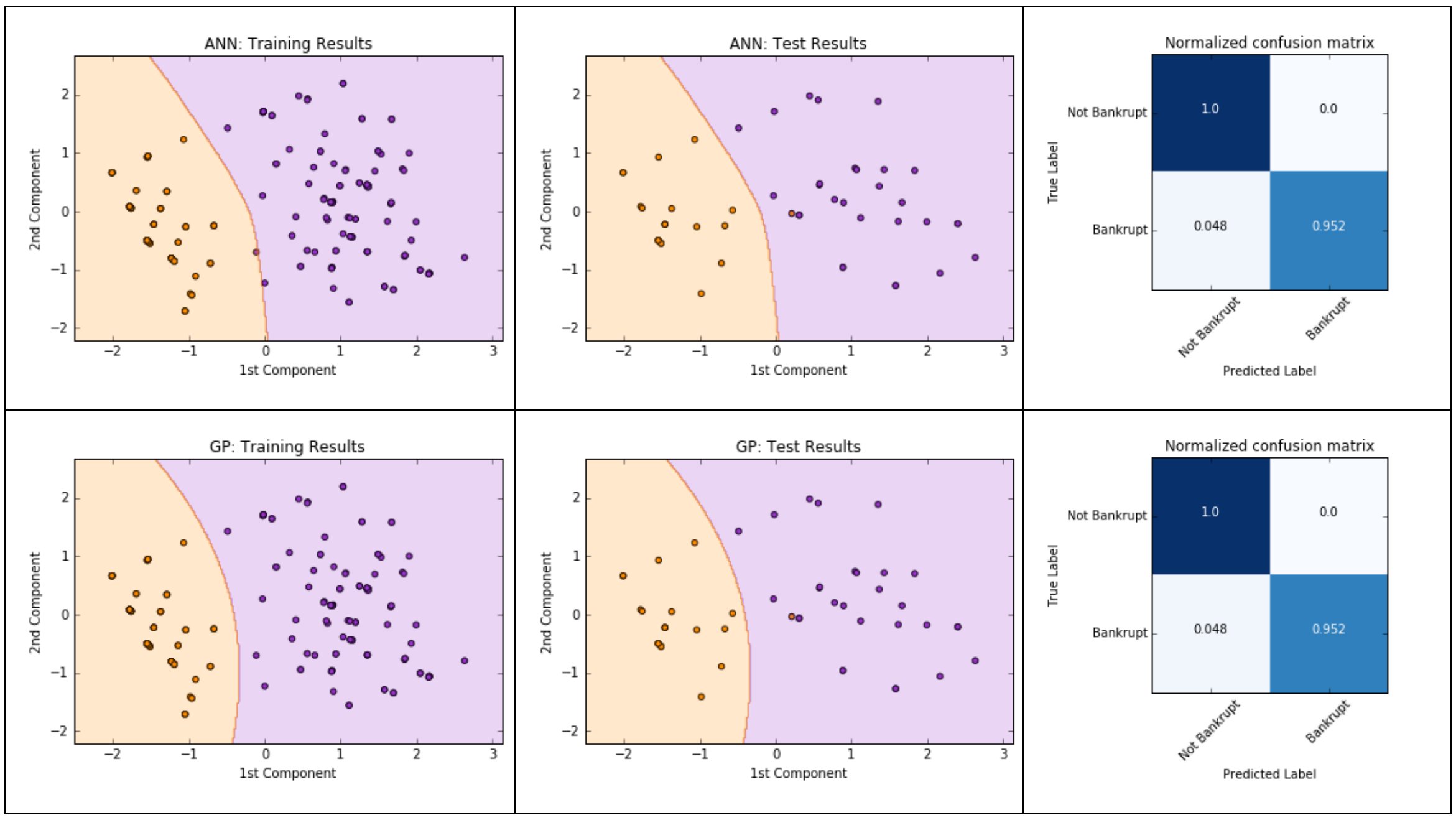

*Figure 13: Decision boundary and confusion matrix of different classifers on PCA transformed features*

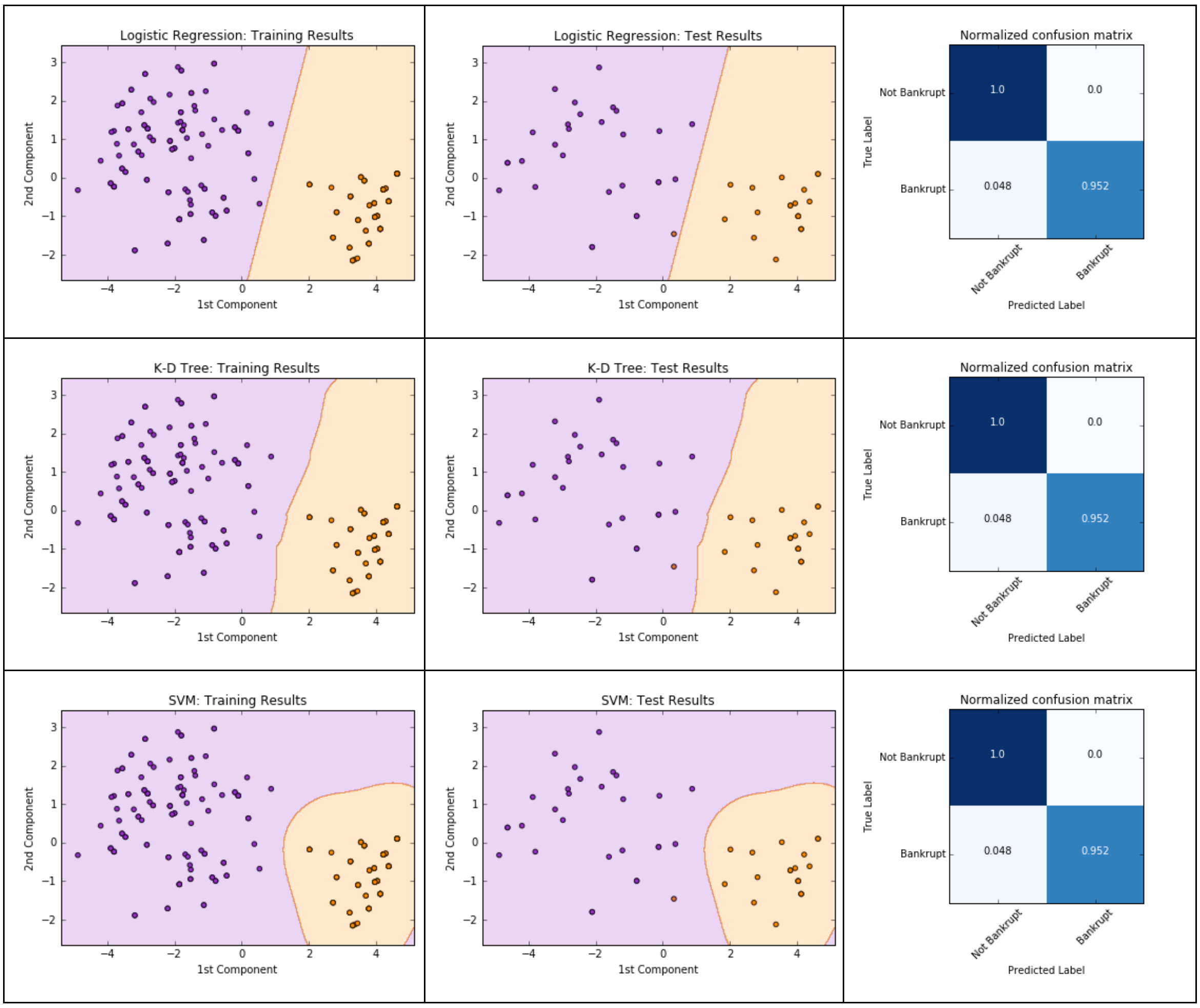

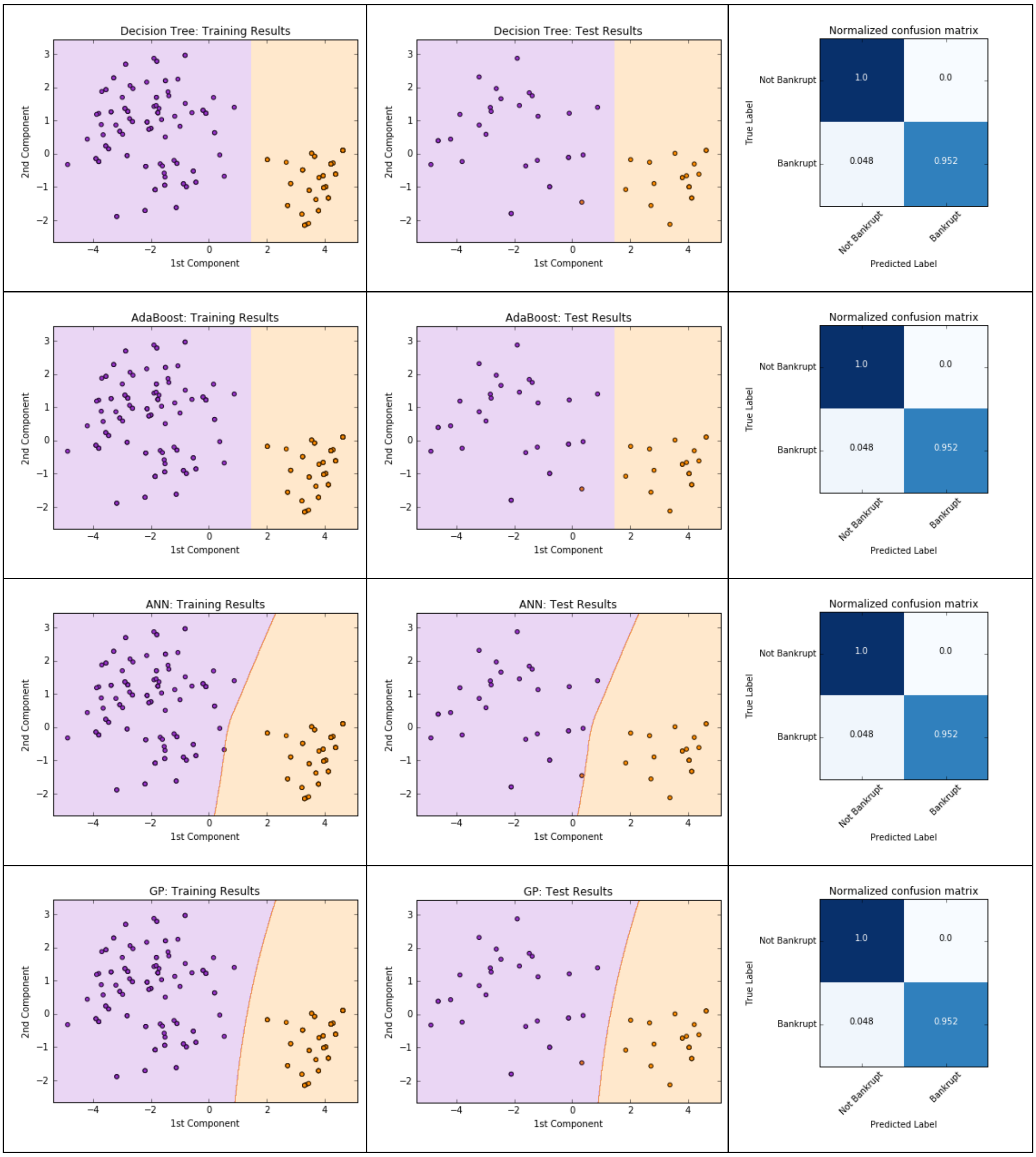

*Figure 14: Decision boundary and confusion matrix of different classifers on LDA transformed features*

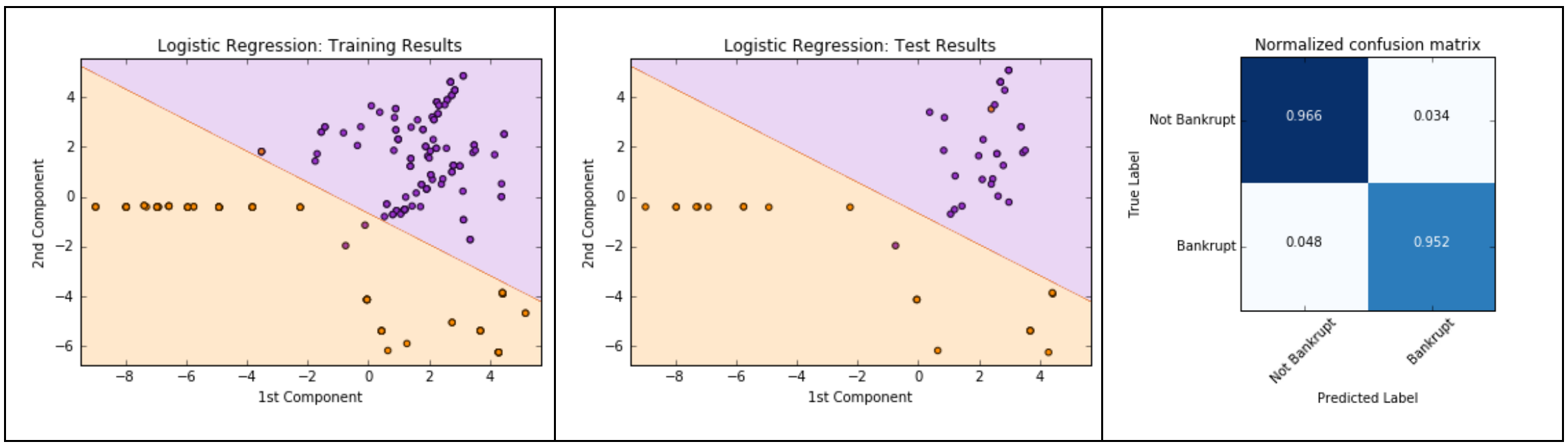

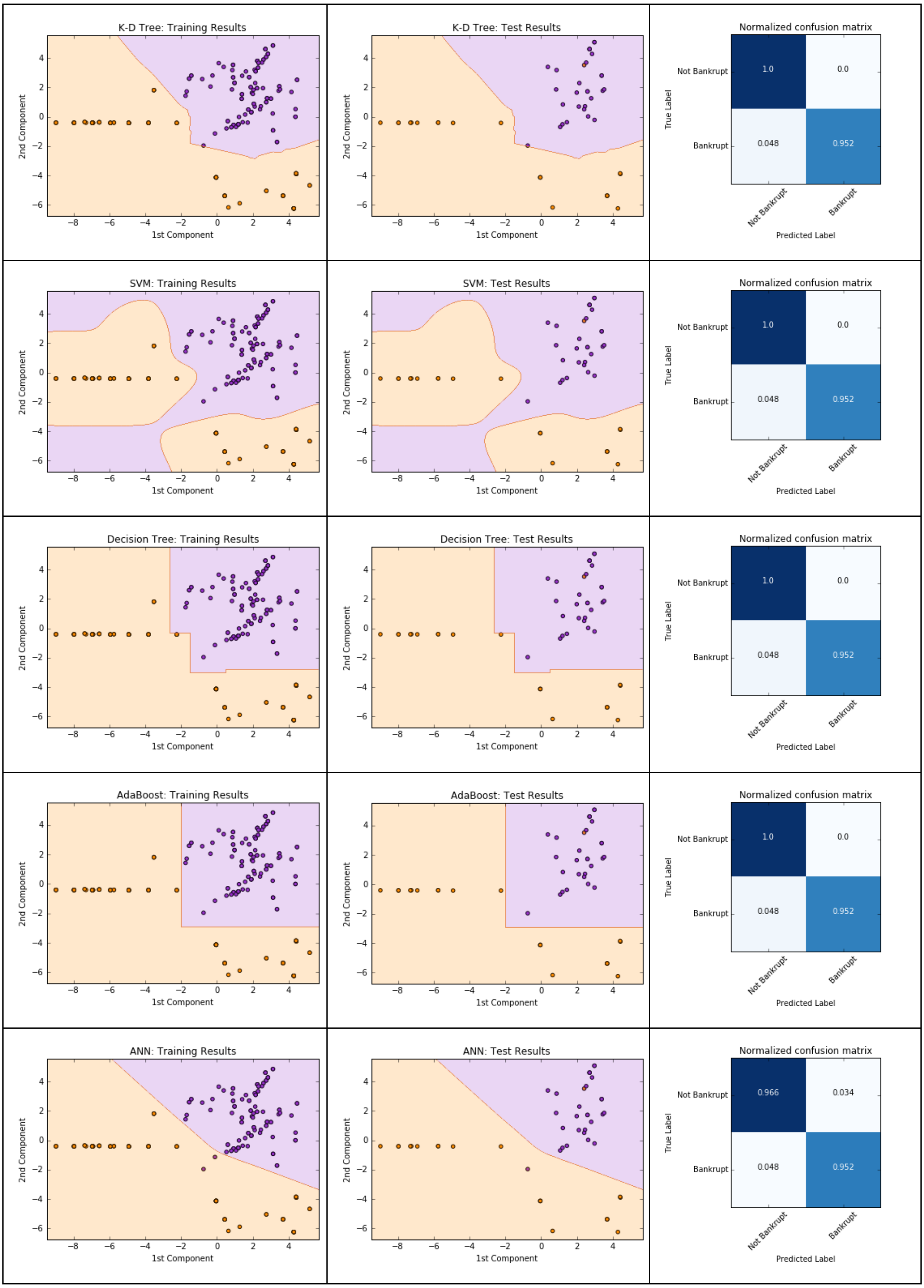

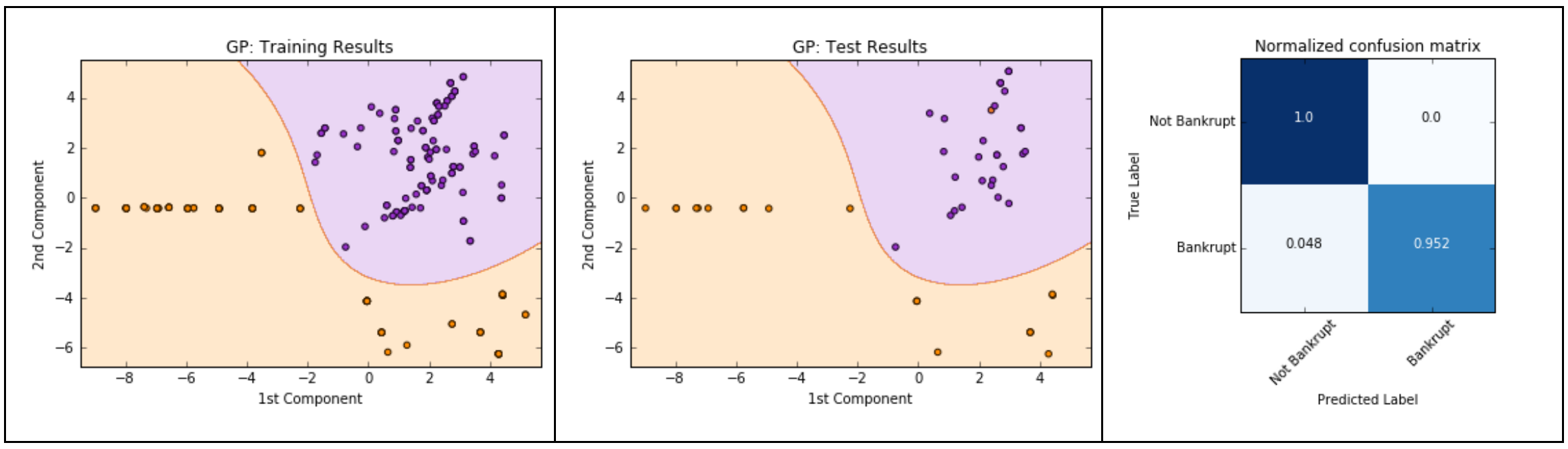

*Figure 15: Decision boundary and confusion matrix of different classifers on ISOMAP transformed features*

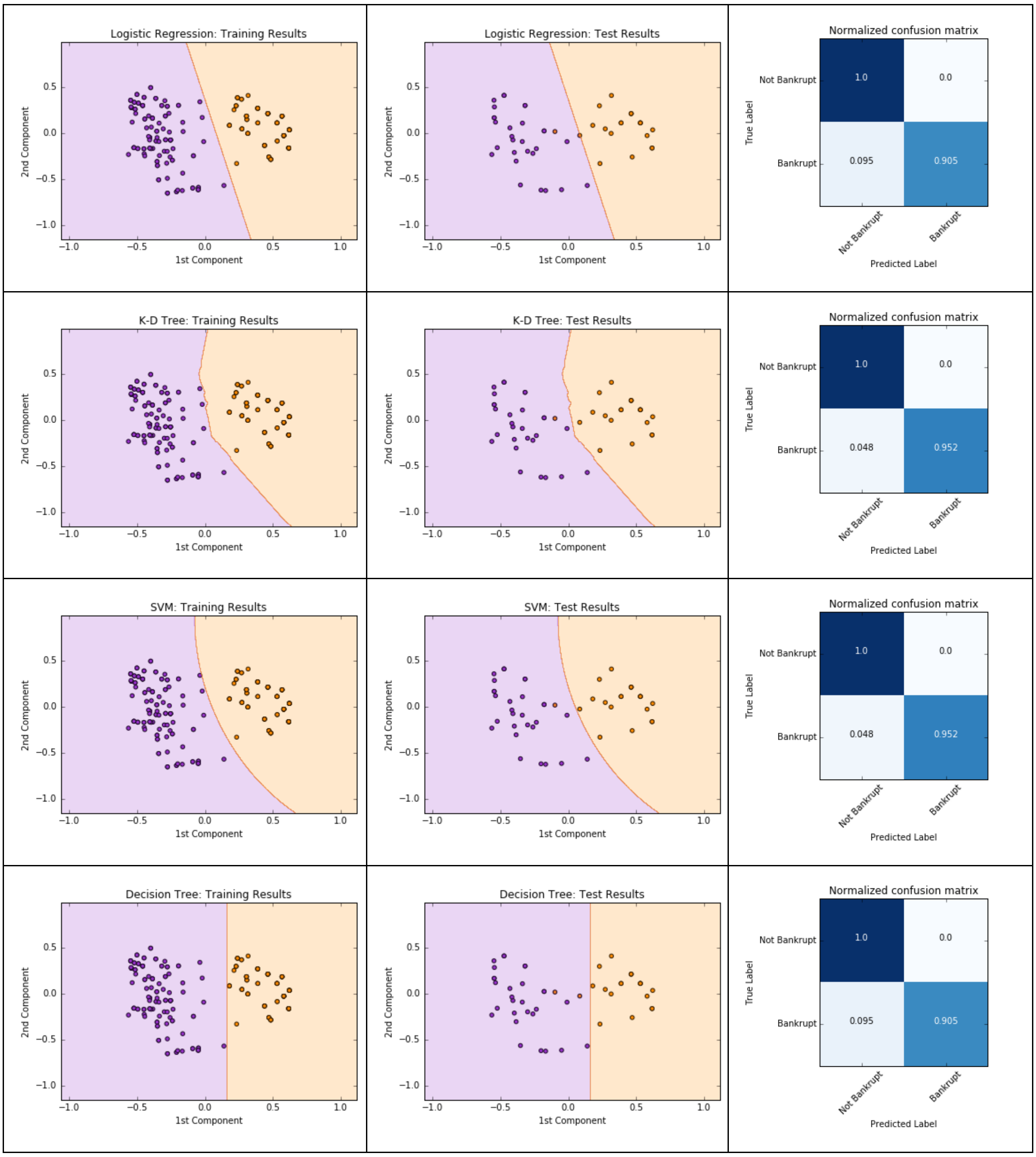

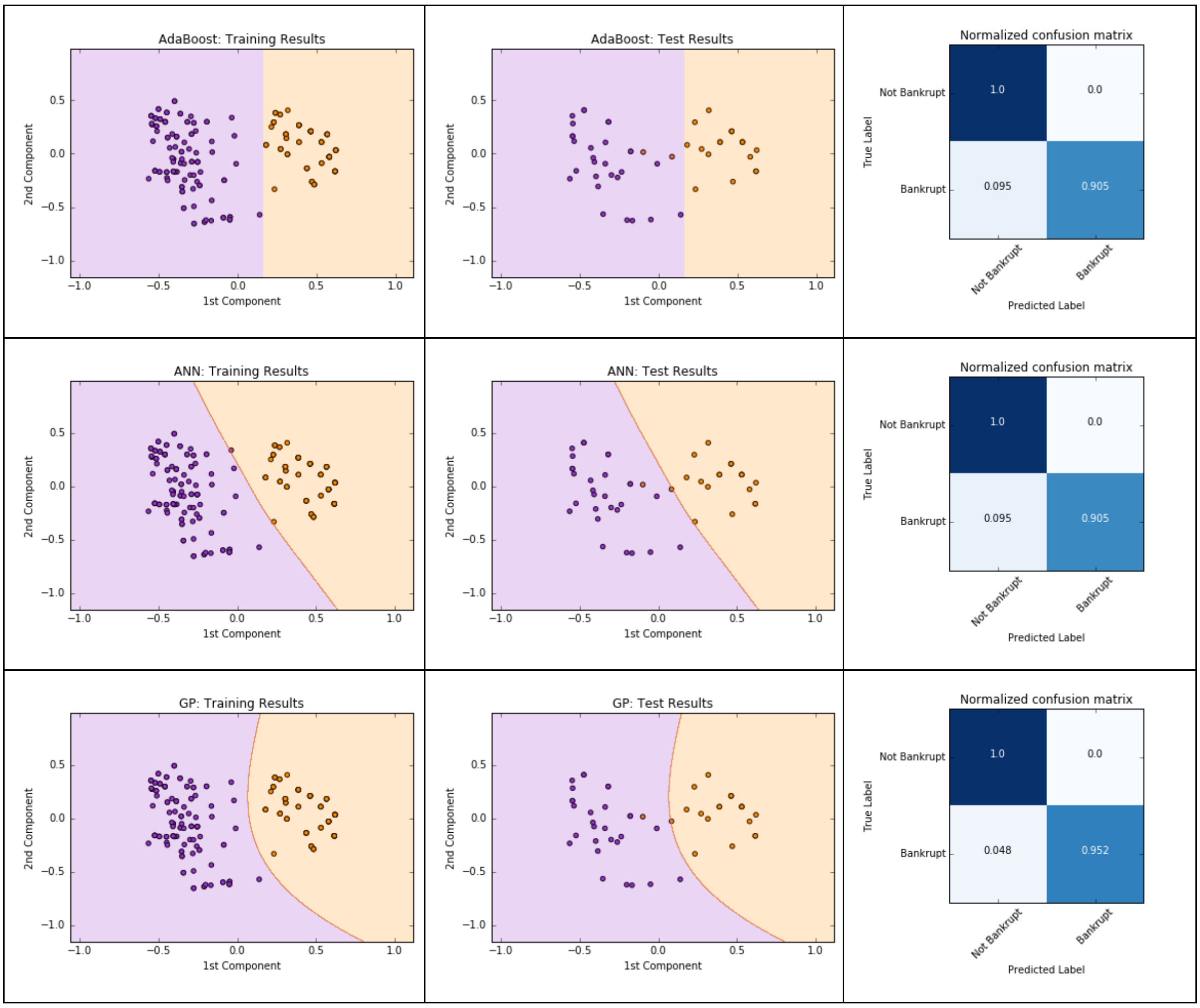

*Figure 16: Decision boundary and confusion matrix of different classifers on Kernel PCA transformed features*

Studying the above confusion matrices reveal that Type-II error (i.e. a company is likely to experience bankruptcy but failed to be detected because it appears "healthy") is more predominant than Type-I error (i.e. flagged company as going bankrupt when in fact it is not). From the perspective of the bank or other loan officers this is unfavourable. Lending money to a company that eventually goes bankrupt can cost the bank more capital than if it foregoes the interest rates on a loan. What most of the classifiers tested above is extremely good at is predicting a company as not going bankrupt when in fact they are financially stable.

Table 1 summarizes the classification accuracy, precision, recall and F1 score for the various binary classification methods using different dimensionality reduction mechanisms. Since the raw input features are discrete (i.e. qualitative measures converted to numerical values), the compute quality measures from the different scenarios are quantized and can be grouped by colours. Green indicates the best performance in that quality measure, yellow is average, and red indicates the worst relative performance. Regardless of the dimensionality reduction method and classification model, the results from all the cases are similar. This suggests that the results are not highly sensitive to the exact model used. Based on Table 1, LDA is the preferred dimensionality reduction method for this dataset because the two-classes are well-separated enough that all classification models, linear or nonlinear, performed equally well, delivering a consistent set of results. Using LDA, the simplest linear separator, namely logistic regression is preferred. In this dataset an overall classification error of 2.0% is achieved. The precision is 100%; reaffirming the fact that if a company is identified as a candidate for insolvency they are almost guaranteed to go bankrupt. While the recall rate of 95.2% suggest that about 5% of the companies experiencing financial distress with pass under the radar and loan officers relying purely on this system for making their decisions will make an error about 5% of the time, which is typically still too high for a bank or financial institution.

*Table 1: Quality control of various machine learning methods on the Korean bankrupty dataset when combined with different dimensionality reduction techniques*

| | | Accuracy | Precision | Recall | F1 Score |
|---|---|---|---|---|---|
| PCA | Logistic Regression | 98.0% | 100.0% | 95.2% | 97.6% |
| | K-D Tree | 98.0% | 100.0% | 95.2% | 97.6% |
| | SVM | 98.0% | 100.0% | 95.2% | 97.6% |
| | Decision Tree | 96.0% | 100.0% | 90.5% | 95.0% |
| | AdaBoost | 96.0% | 100.0% | 90.5% | 95.0% |
| | ANN | 98.0% | 100.0% | 95.2% | 97.6% |
| | GP | 98.0% | 100.0% | 95.2% | 97.6% |
| LDA | Logistic | 98.0% | 100.0% | 95.2% | 97.6% |
| | Regression | | | | |
| | K-D Tree | 98.0% | 100.0% | 95.2% | 97.6% |
| | SVM | 98.0% | 100.0% | 95.2% | 97.6% |
| | Decision Tree | 98.0% | 100.0% | 95.2% | 97.6% |
| | AdaBoost | 98.0% | 100.0% | 95.2% | 97.6% |
| | ANN | 98.0% | 100.0% | 95.2% | 97.6% |
| | GP | 98.0% | 100.0% | 95.2% | 97.6% |
| ISOMAP | Logistic Regression | 96.0% | 95.2% | 95.2% | 95.2% |
| | K-D Tree | 98.0% | 100.0% | 95.2% | 97.6% |
| | SVM | 98.0% | 100.0% | 95.2% | 97.6% |
| | Decision Tree | 98.0% | 100.0% | 95.2% | 97.6% |
| | AdaBoost | 98.0% | 100.0% | 95.2% | 97.6% |
| | ANN | 96.0% | 95.2% | 95.2% | 95.2% |
| | GP | 98.0% | 100.0% | 95.2% | 97.6% |
| Kernel PCA | Logistic Regression | 96.0% | 100.0% | 90.5% | 95.0% |
| | K-D Tree | 98.0% | 100.0% | 95.2% | 97.6% |
| | SVM | 98.0% | 100.0% | 95.2% | 97.6% |
| | Decision Tree | 96.0% | 100.0% | 90.5% | 95.0% |
| | AdaBoost | 96.0% | 100.0% | 90.5% | 95.0% |
| | ANN | 96.0% | 100.0% | 90.5% | 95.0% |
| | GP | 98.0% | 100.0% | 95.2% | 97.6% |

Different classifiers have different strengths. To analyse the insolvency situation in this Korean market further, the decision tree model is used. As mentioned in Section 2.2.4, one of the benefits of using decision trees is the transparency it provides when tracing back the decision-making process. Take the ISOMAP projected 2D features as an example, after the training is complete, the decision tree can be visualized as shown in Figure 17. Beginning at the first node on the top with 200 samples and a Gini index of 0.49, if the 2$^{nd}$ component from the projected features is less than -0.33 traverse to the left branch, else traverse to the right branch. This process repeats until the Gini index is zero at a node, where each node is either going to be classified as bankrupt or not bankrupt. As one can imagine, typically the more features and the more complicated the decision boundary the larger the tree structure, therefore the more difficult for a human analyst to interpret the results.

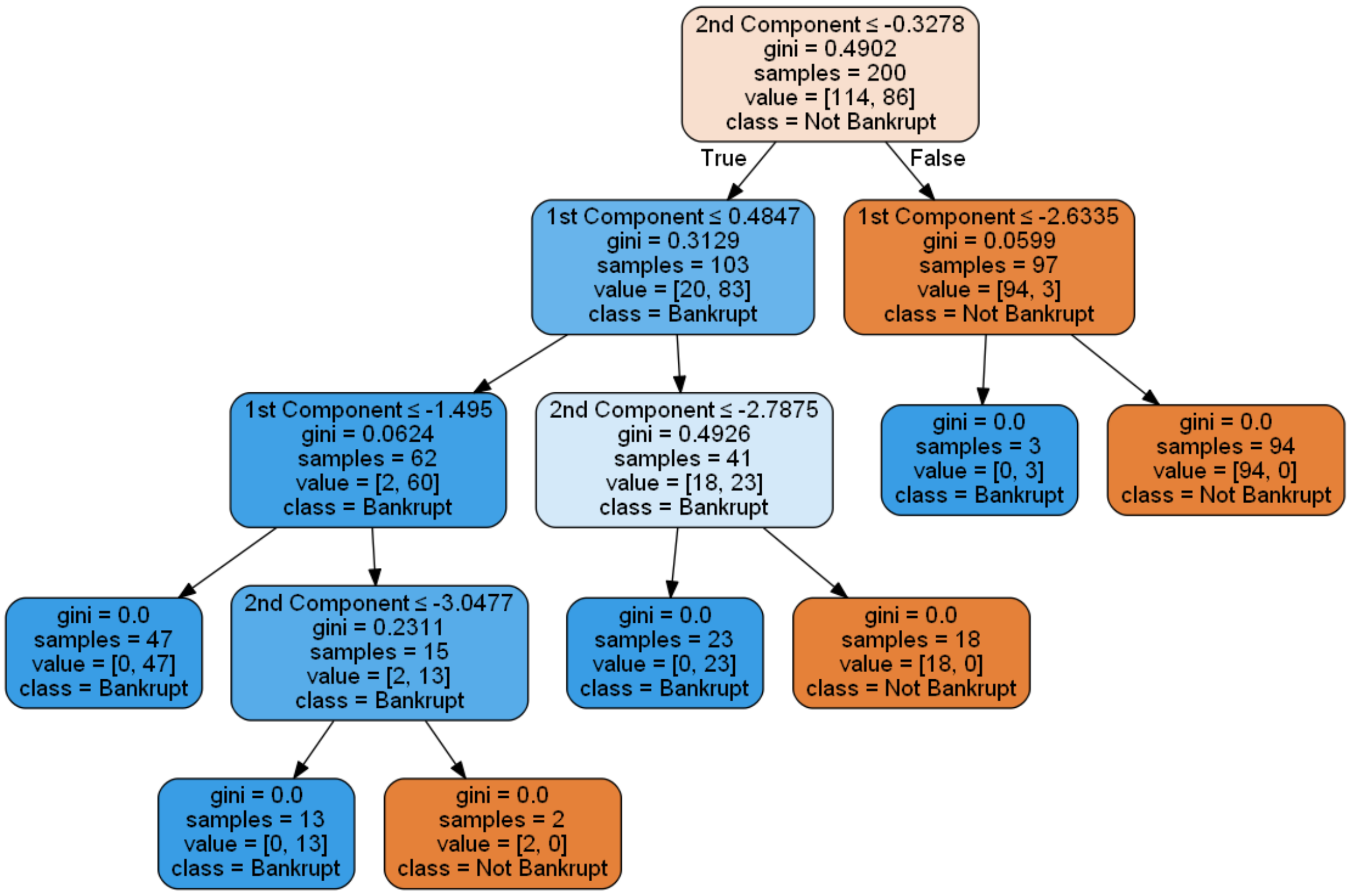

*Figure 17: Trained decision tree on the ISOMAP transformed features in two-dimensional space*

Although using dimensionality reduction can typically reduce the complexity of the problem, the physical meaning of the feature space is also partly lost due to the projection, for instance the 1st and 2nd component does not have much meaning to a financial expert. One possible solution is to study the projection function and indirectly relate the tree structure to the originally measured features. For example, when using PCA the weights for each feature contributing to the projection can be extracted from the eigenvectors. For this particular Korean dataset the weights for the first and second component are given in Table 2. It can be observed that the 1st component is dominated by the financial flexibility, credibility, and competitiveness of the firm, and the 2nd component is mostly expressing the industrial risk, management risk, and operating risk of the firm.

*Table 2: Weights of the original six-dimensional qualitative features on the projected two-dimensional principal components*

| | Industrial Risk | Management Risk | Financial Flexibility | Credibility | Competitiveness | Operating Risk |
|---|---|---|---|---|---|---|
| 1st Component | 0.231 | 0.320 | 0.466 | 0.472 | 0.585 | 0.250 |
| 2nd Component | -0.582 | -0.307 | 0.268 | 0.284 | 0.214 | -0.607 |

Alternatively, the decision tree classifier can be applied to the original six-dimensional dataset to directly learn the relationship between each qualitative measure and the likelihood of the company experiencing financial distress (Figure 18). In this specific case, the resulting tree structure is relatively simple (it is deeper but less wide than the tree in Figure 17). From Figure 18, it can be seen that only four of the six qualitative measures mattered for this particular training set. More specifically, in the order of importance they are competitiveness, credibility, financial flexibility, and industrial risk. Stepping through this decision tree in detail, the following can be said about these two hundred Korean manufacturing and service companies:

1. The firm's competitiveness has the strongest impact on insolvency. If the company receives a negative competitive score it will likely go bankrupt regardless of its score in other attributes.
2. If the firm has an average or high competitive level and good credibility, then this firm will unlikely face bankruptcy.

3. But if this competitive firm has poor credibility then it better be financially flexible or be in an industry where the industrial risk is low for them to stay afloat.

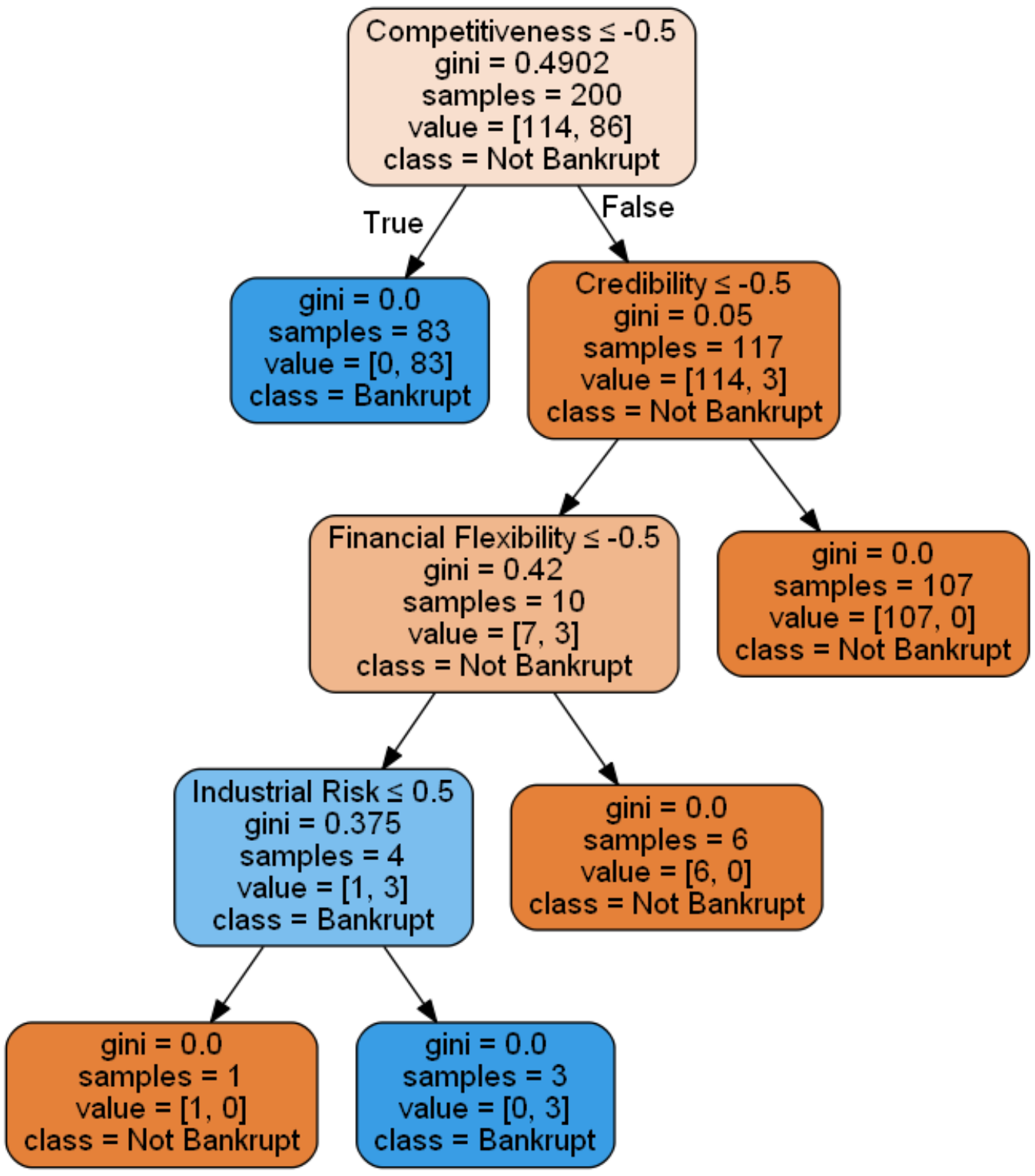

*Figure 18: Trained decision tree on the raw features*

Compared to the original authors' results (Kim & Han, 2003), the presented solution in this thesis showed better classification accuracy. In Kim and Han (2003), the overall binary classification accuracy of their genetic algorithm, inductive learning, and neural network are reported to be 94.0%, 89.7%, and 90.3%, respectively. In this thesis, better classification accuracy (i.e. 98%) is achieved using various classification models. This improvement can be attributed to the smaller sample size (in reference to the original paper) and the use of dimensionality reduction techniques. In the article from Kim and Han (2003), all their data mining technique are applied to the original feature space, which as demonstrated in this thesis, contains noise and irrelevant information for bankruptcy prediction. Therefore, it is reasonable to eliminate some dimensions in the dataset that are less informative to reduce the complexity of the problem.

## 5.2 Dataset 2: Polish Corporate Bankruptcy

### 5.2.1 Visualization of the Data

Following the same approach as 5.1.1 the quantitative financial factors are first scaled to be between -1.0 and +1.0. Unlike Dataset 1, Dataset 2 is incomplete in the sense that there are missing attributes for some companies. Instead of simply removing these entries entirely from the dataset, their missing values are imputed from the remaining companies as the median. This way the number of samples remains the same and the imputed features will have little influence on the decision boundary (Jereza, et al., 2010). Furthermore, there are significantly more data that are labelled as "not bankrupt" (the ratio of not bankrupt to bankrupt companies in this database is 13:1). This class imbalance can have a significant impact on some classification algorithms. Therefore the features labelled as "bankrupt" are up-sampled using the Synthetic Minority Over-sampling Technique (Chawla, Bowyer, Hall, & Kegelmeyer, 2002).

Dataset 2 with quantitative financial measures is more complex than Dataset 1, using the same dimensionality reduction methods for mapping features into three-dimensional space revealed that the bankrupt and non-bankrupt companies are very similar in a low dimensional space (Figure 19). Therefore, a visual assessment of the classification boundary is not possible and the assessment has to be based on other numerical techniques.

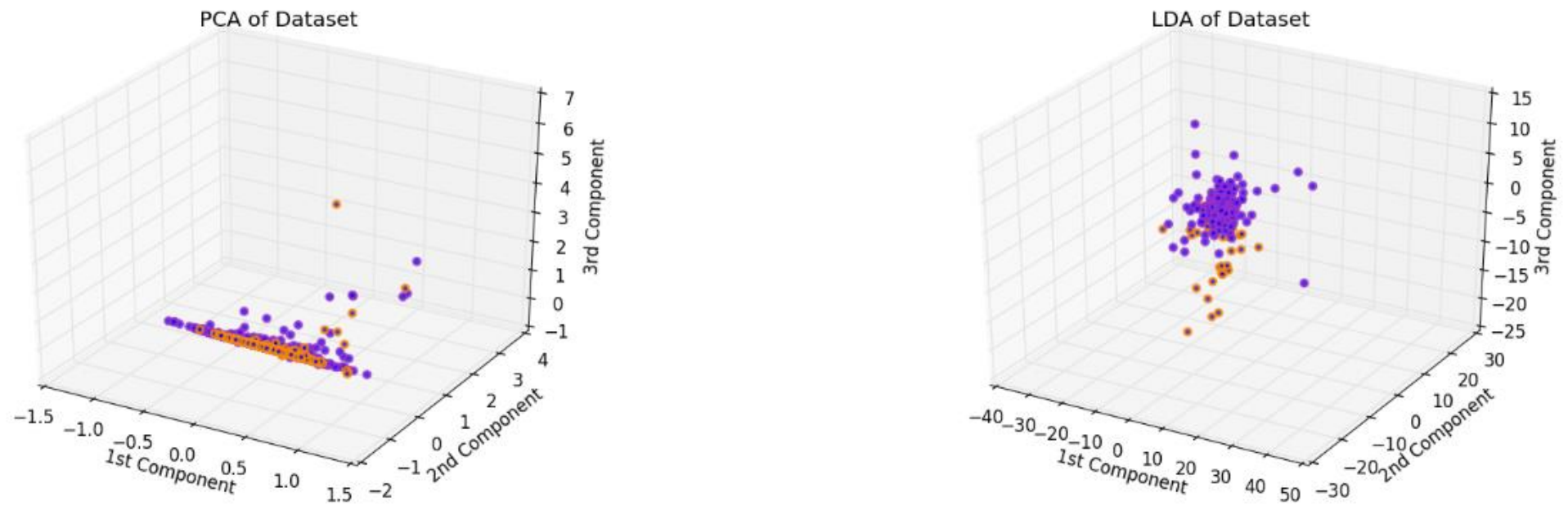

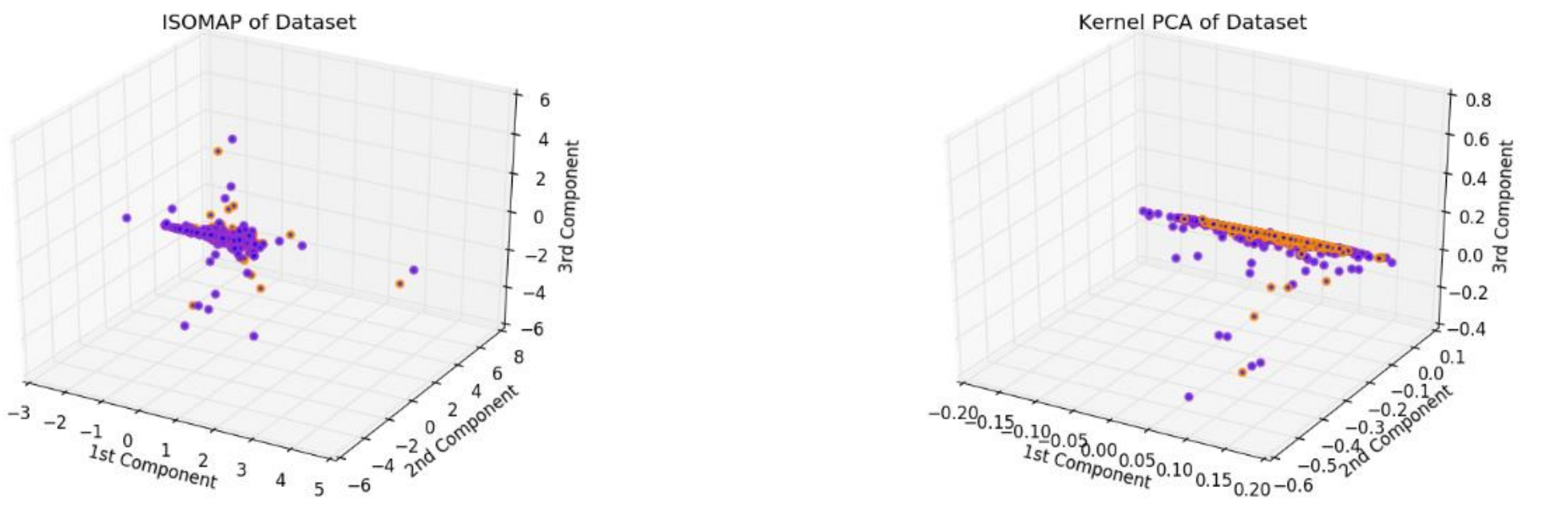

*Figure 19: Three-dimensional visualization of the Polish bankrupty dataset*

To choose a suitable number of components without visualizing the point distribution the PCA components can be plotted in descending order of information content as shown in Figure 20. Visually, 50% of the variance can be summarized by the first couple components. Principal components 31 to 64 together accounts for only 1% of variance in the data and is assumed to be highly contaminated by noise. Therefore, the various classification algorithms is applied to only the 30 largest principal components, which should capture 99% of the information while reducing the dimensionality of the problem by more than half.

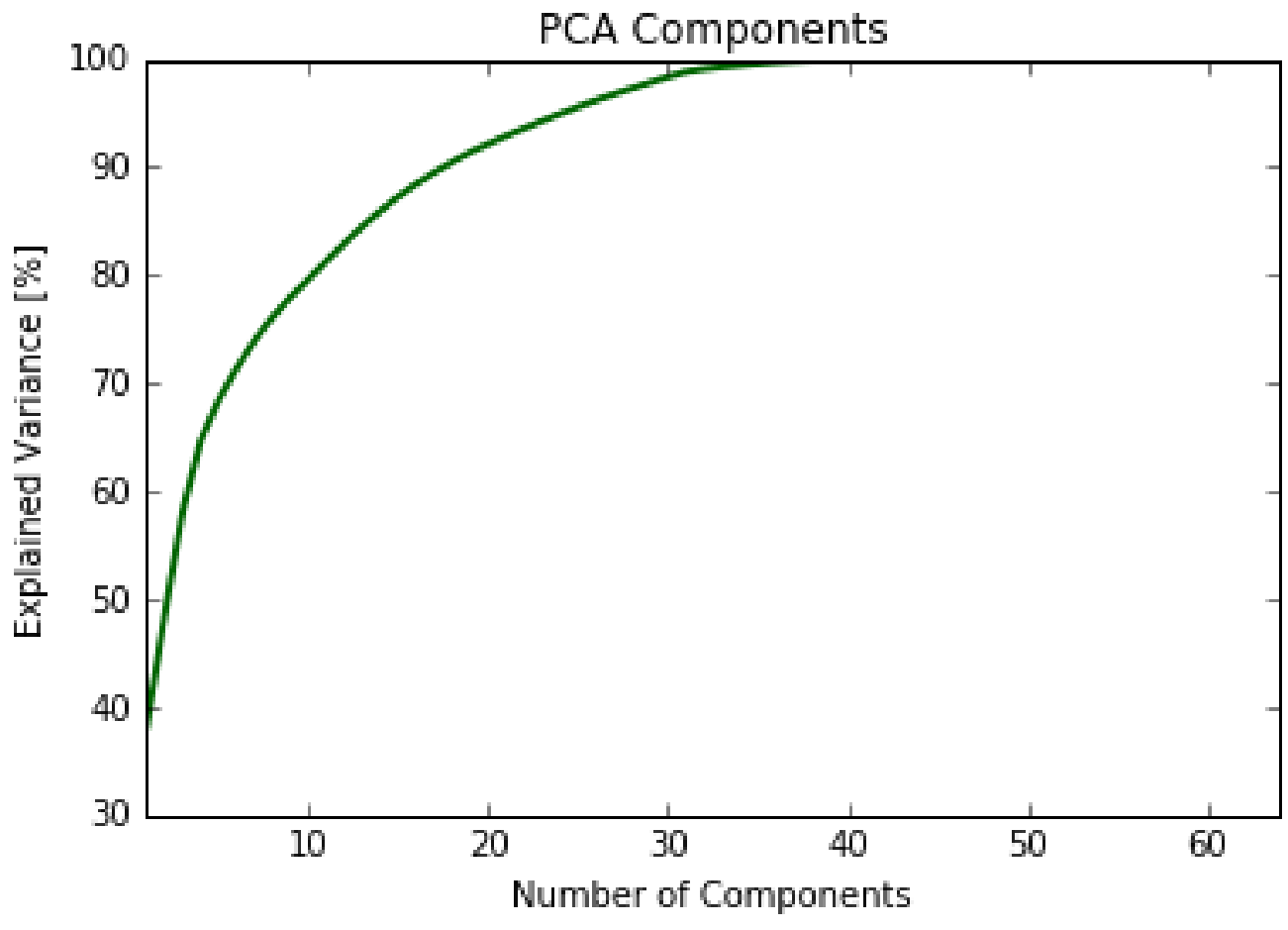

*Figure 20: The accumulated percentage of variance in the principal components*

### 5.2.2 Binary Classification

All hyper-parameters are tuned using a 10-fold cross-validation on the training data (e.g. for K-D Tree the optimal neighbourhood size was found to be six in this case). This allows the “best” model to be selected in an objective fashion using the data itself. Figure 21 shows the 10-fold ROC curve of the training data and the confusion matrix of the testing data. Between the individual folds, ANN and GP displayed the largest discrepancies in their ROC curves. K-D tree and decision tree exhibit the most preferable shape for their ROC curves. Even though their ROC curves look similar, their confusion matrices are quite a contrast; K-D tree can detect bankrupt companies with a 57.0% accuracy while the trained decision tree has only a 11.6% accuracy.

<table>
<tr>
<td>Logistic Regression</td>
<td>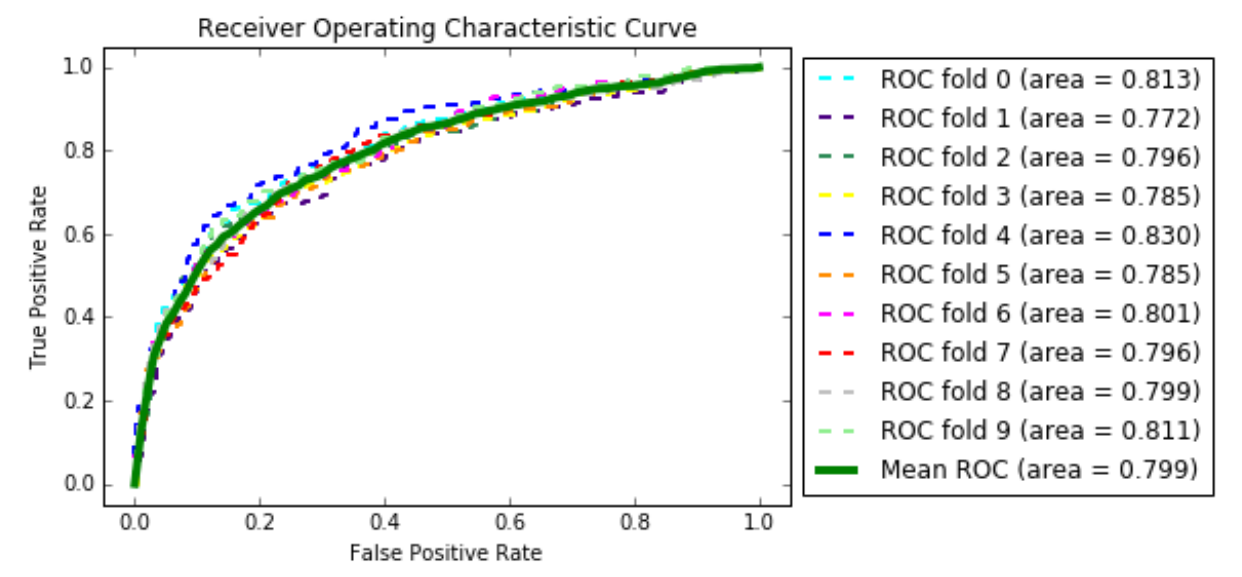
</td>
<td>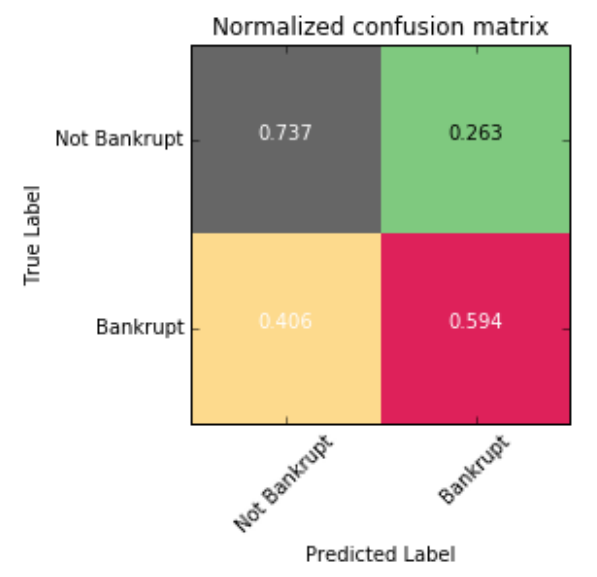
</td>
</tr>
<tr>
<td>K-D Tree</td>
<td>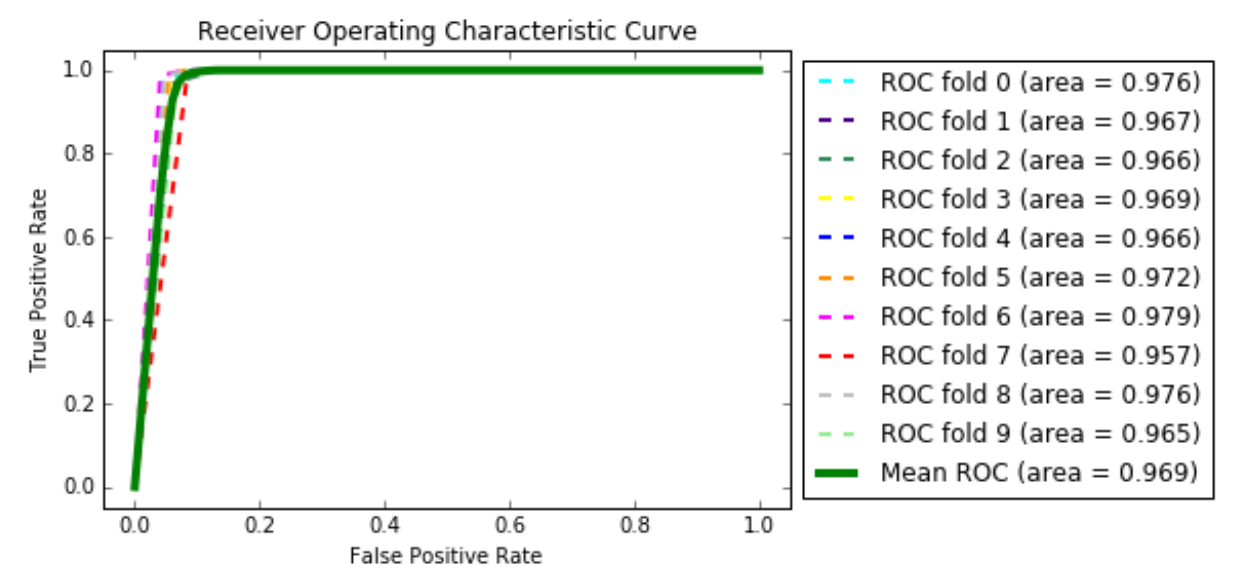
</td>
<td>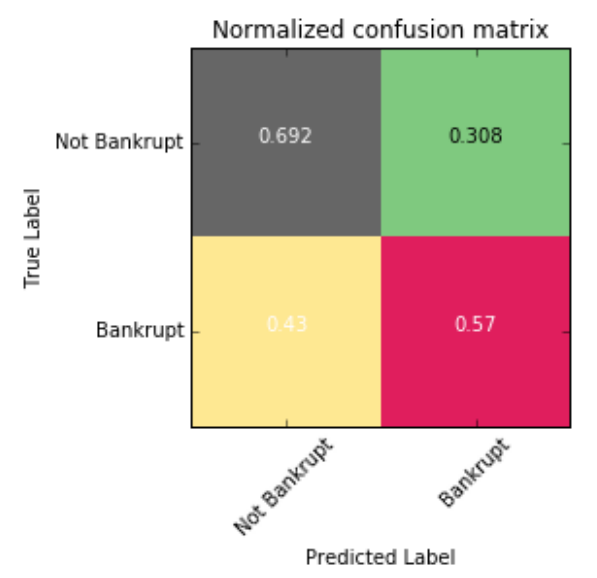
</td>
</tr>
<tr>
<td>SVM</td>
<td>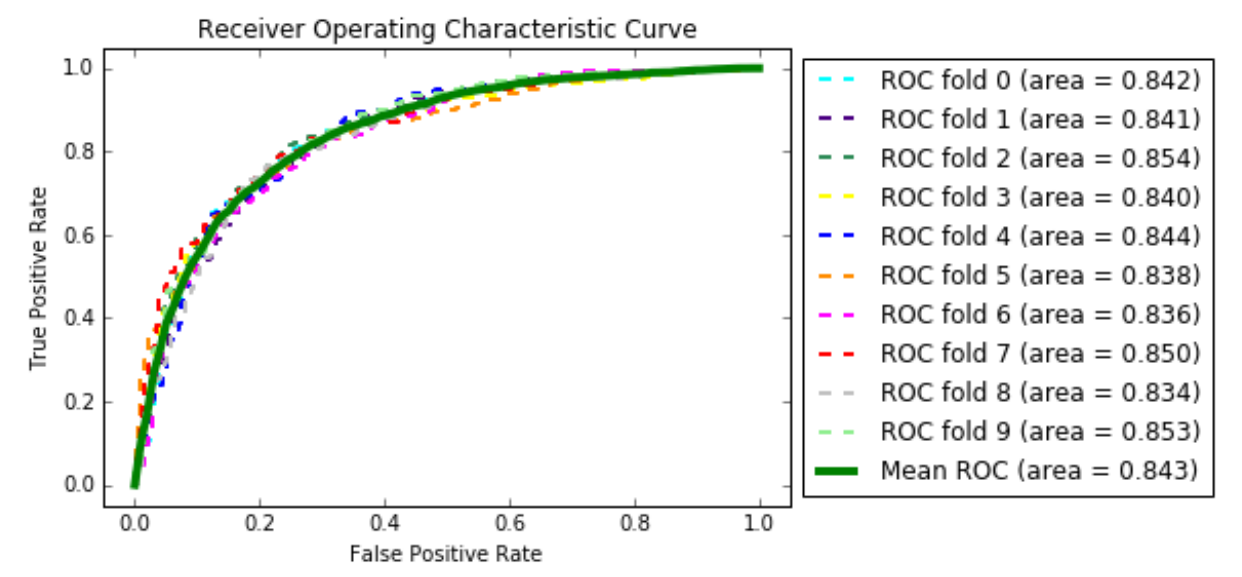
</td>
<td>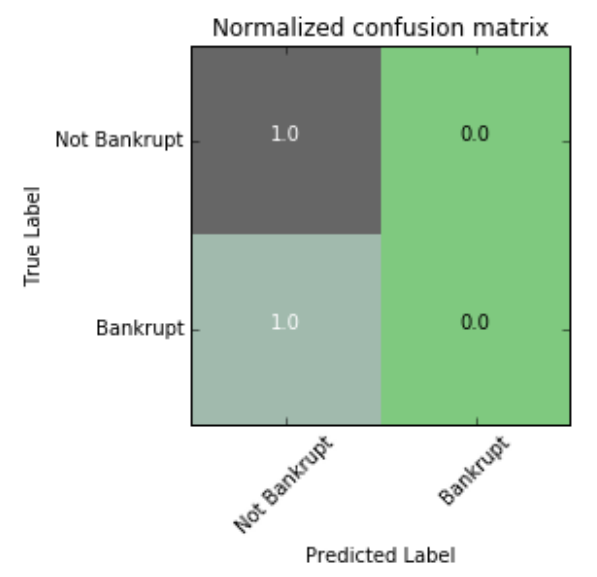
</td>
</tr>
</table>

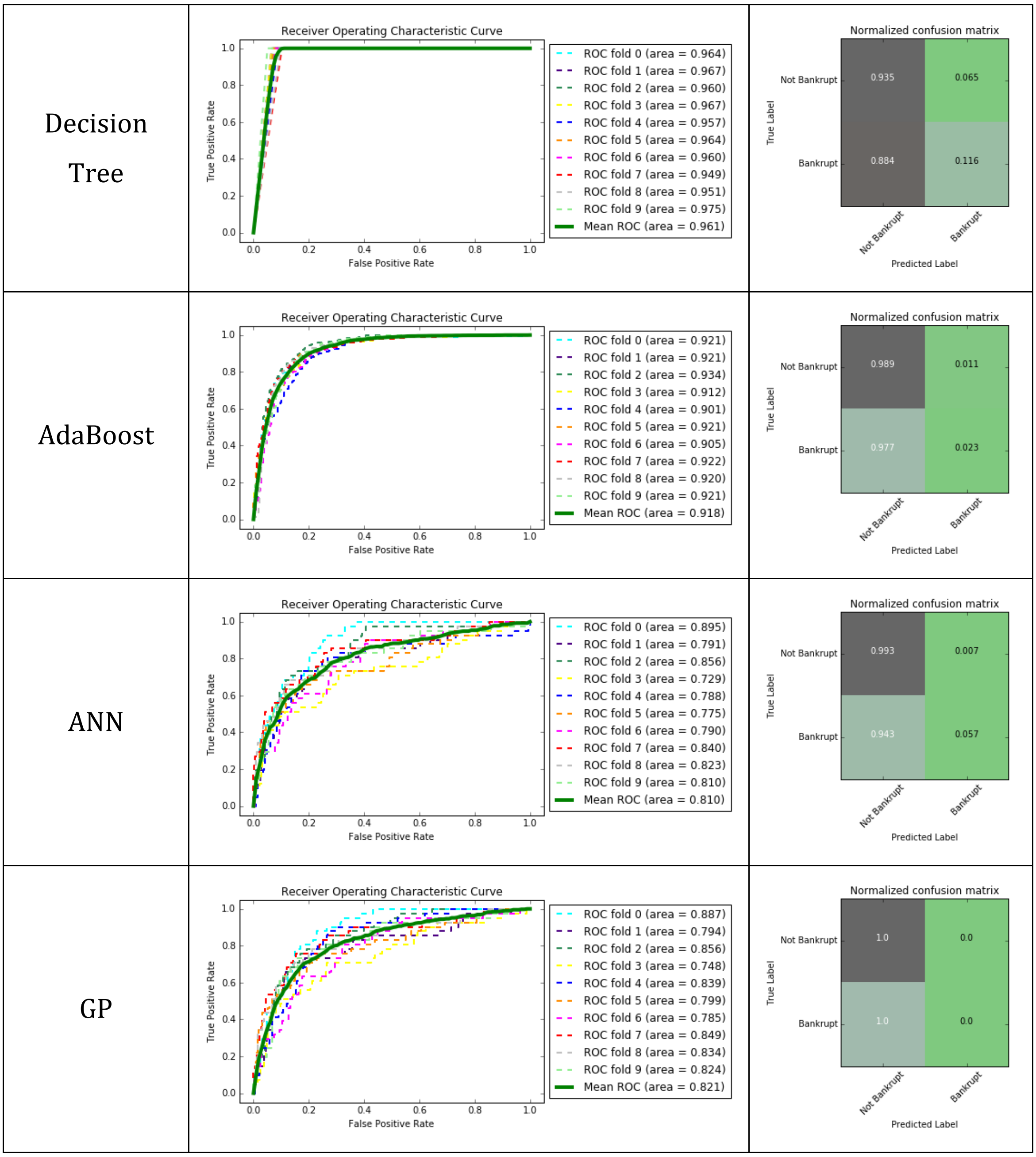

*Figure 21: ROC curve and confusion matrix from applying different machine learning methods on the Polish bankruptcy dataset*

Quality control measures of the different machine learning models are summarized in Table 3. Besides accuracy, precision, recall, and F1 score, the mean area under a ROC curve (AUC) is also reported to allow for easier comparison with the results from the original article by Zięba et al. (2016). Some classifiers like logistic regression and SVM are also tested in Zięba et al. (2016), and overall the mean AUC obtained in this thesis is comparable. For instance, the AUC for logistic regression reported in this thesis is 79.9% while in the original article it is 63.2%. Considering that the classifier is the same, the improvement in the performance can likely be attributed to the data scaling, dimensionality reduction, and model selection steps.

Not only are the AUC scores in this thesis similar to the original article, the highest AUC of 96.9% exceeded all the methods reported in Zięba et al. (2016). But it should be noted that AUC alone is insufficient to describe the actual performance of the classifier. Many classifiers in Table 3 have a high AUC (i.e. greater than 90%) but their accuracy, precision, recall, and F1 score can vary significantly. For instance, AdaBoost has a higher AUC than ANN (91.8% compared to 81.0%), but ANN exhibits better accuracy, precision, recall, and F1 score. When analysing all the quality measures together it can be seen that the nonlinear classifiers tend to have less consistent measures. For example, SVM reported good AUC and accuracy (i.e. 84.3% and 92.6%, respectively) but the precision, recall, and F1 score is practically zero because the classifier is predicting all data as belonging to the non-bankrupt class as indicated by the confusion matrix. This is likely a result of the curse of dimensionality where these higher-order nonlinear classifiers are over-fitting due to the limited amount of training data for such a high dimensional problem. Typically, with increasing number of features (i.e. dimensions) there needs to be an exponential growth in the number of data to maintain the same density (Verleysen & François, 2005). Although through dimensionality reduction the number of features are significantly reduced compared to the original paper, 5000 data points for a 30-dimensional space is still rather low.

After analysing other quality control factors it appears that due to the limited amount of data and the high number of features, simpler classifiers such as K-D tree and logistic regression (despite having the lowest AUC among the classifiers tested, it has a better average quality control measure) are actually able to generalize better to new data than more sophisticated machine learning algorithms. On the test data, their performances are comparable, with logistic regression performing slightly better and require significantly less storage to make faster predictions.  Thus, it is the preferred machine learning model for this dataset because is it well balanced.

*Table 3: Quality control of different machine learning methods on the Polish bankruptcy dataset*

| | Accuracy | Precision | Recall | F1 Score | AUC |
|---|---|---|---|---|---|
| Logistic Regression | 66.4% | 70.4% | 59.4% | 64.4% | 79.9% |
| K-D Tree | 62.9% | 66.1% | 57.0% | 61.2% | 96.9% |
| SVM | 92.6% | 0.0% | 0.0% | 0.0% | 84.3% |
| Decision Tree | 51.5% | 65.4% | 11.6% | 19.7% | 96.1% |
| AdaBoost | 91.8% | 14.3% | 2.3% | 4.0% | 91.8% |
| ANN | 92.4% | 38.5% | 5.7% | 10.0% | 81.0% |
| GP | 92.6% | 0.0 | 0.0 | 0.0 | 82.1% |

# 6 Conclusion and Recommendations

Business intelligence is a rapidly evolving field carrying tremendous potential for improving efficiency and competitiveness of corporations across the globe. Data mining and machine learning are useful tools in this field that can extract valueable information from big data to aid business strategies. This thesis investigated the adoption of several popular modern day machine learning algorithms for forecasting business failures, i.e. corporate bankruptcy. This information can help governments, investors, managers, and other stakeholders make intelligent economic decisions to avoid financial losses.

Two very different datasets of companies in the manufacturing sector were analysed and compared to other researchers' results. The first dataset uses six qualitative measures to describe the business situation for companies in Korea. The second dataset uses 64 quantitative financial features for assessing the likelihood of corporate insolvency in Poland. Results from this thesis indicated that all machine learning algorithms applied to Dataset 1 yielded superior performance when compared to Dataset 2. This can be attributed to the fact that these handcrafted features from experts' advice are more expressive than financial ratios. Despite only asking the financial experts six multiple choice questions about each company, the collected data showed a more distinct separation between the two classes. Actually, the decision tree classifier only required four out of the six questions to detect companies with high chances of insolvency with better than 90% accuracy. This suggested that in the future, fewer questions can be asked, collected, and stored. The informativeness of these qualitative measures is advantageous because even after projecting the data onto a lower dimensional subspace for visualization a clear separation between the two clusters can still be perceived, making it more intuitive to optimize, compare, and analyse the classification methods.

In Dataset 2, even with 64 financial ratios and an order of magnitude more data, it was difficult for the machine learning algorithms to classify companies as going to bankrupt or not with robustness. However, collecting experts' advice is quite expensive compared to using financial ratios that are readily available in most companies' balance sheets and income statements. Considering that collecting a large number of financial ratios is typically less expensive than capturing experts' opinions, it is recommended that more data to be collected for training the classifier to combat the curse of dimensionality and lack of expressiveness of individual features. Quality control should also be based on multiple performance measures rather than relying on a single score. As demonstrated in this thesis, when analysing and comparing machine learning techniques using a single quality control metric, the conclusion can easily be biased.

In general, machine learning is a powerful toolbox for financial analysts to make predictions and discover patterns in the data with rigour. Many different models and validation techniques exist to aid data mining and decision making. It is difficult to determine if any machine learning technique is superior to others. In fact, as a data scientist and/or financial expert, it is perhaps more beneficial to harvest the strengths of different methods and combine them to make better business judgements.

# 8 Appendices

# APPENDIX 1: KOREAN BANKRUPTCY DATASET FEATURES

The following is a detailed description of the six qualitative features from the Korean bankruptcy dataset:

| *Risk Factor* | *Risk Component* |
|---|---|
| Industry Risk | Government policies and International agreements<br>Cyclicality<br>Degree of competition<br>The price and stability of market supply<br>The size and growth of market demand<br>The sensitivity to changes in macroeconomic factors<br>Domestic and international competitive power<br>Product Life Cycle<br>IR |
| Management Risk | Ability and competence of management<br>Stability of management<br>The relationship between management/owner<br>Human resources management<br>Growth process/business performance<br>Short and long term business planning, achievement and feasibility |
| Financial Flexibility | Direct financing<br>Indirect financing<br>Other financing (affiliates, owner, third parties) |
| Credibility | Credit history<br>The reliability of information<br>The relationship with financial institutes |
| Competitiveness | Market position<br>The level of core capacities<br>Differentiated strategy |
| Operating Risk | The stability and diversity of procurement<br>The stability of transaction<br>The efficiency of production<br>The prospects for demand for product and service<br>Sales diversification<br>Sales price and settlement condition<br>Collection of A/R<br>Effectiveness of sale network |

# APPENDIX 2: POLISH BANKRUPTCY DATASET FEATURES

The following is a complete list of the 64 quantitative features from the Polish bankruptcy dataset:

| | |
|---|---|
| 1 | net profit / total assets |
| 2 | total liabilities / total assets |
| 3 | working capital / total assets |
| 4 | current assets / short-term liabilities |
| 5 | [(cash + short-term securities + receivables - short-term liabilities) / (operating expenses - depreciation)] * 365 |
| 6 | retained earnings / total assets |
| 7 | EBIT / total assets |
| 8 | book value of equity / total liabilities |
| 9 | sales / total assets |
| 10 | equity / total assets |
| 11 | (gross profit + extraordinary items + financial expenses) / total assets |
| 12 | gross profit / short-term liabilities |
| 13 | (gross profit + depreciation) / sales |
| 14 | (gross profit + interest) / total assets |
| 15 | (total liabilities * 365) / (gross profit + depreciation) |
| 16 | (gross profit + depreciation) / total liabilities |
| 17 | total assets / total liabilities |
| 18 | gross profit / total assets |
| 19 | gross profit / sales |
| 20 | (inventory * 365) / sales |
| 21 | sales (n) / sales (n-1) |
| 22 | profit on operating activities / total assets |
| 23 | net profit / sales |
| 24 | gross profit (in 3 years) / total assets |
| 25 | (equity - share capital) / total assets |
| 26 | (net profit + depreciation) / total liabilities |
| 27 | profit on operating activities / financial expenses |
| 28 | working capital / fixed assets |
| 29 | logarithm of total assets |
| 30 | (total liabilities - cash) / sales |
| 31 | (gross profit + interest) / sales |
| 32 | (current liabilities * 365) / cost of products sold |
| 33 | operating expenses / short-term liabilities |
| 34 | operating expenses / total liabilities |
| 35 | profit on sales / total assets |
| 36 | total sales / total assets |
| 37 | (current assets - inventories) / long-term liabilities |
| 38 | constant capital / total assets |
| 39 | profit on sales / sales |
| 40 | (current assets - inventory - receivables) / short-term liabilities |
| 41 | total liabilities / ((profit on operating activities + depreciation) * (12/365)) |
| 42 | profit on operating activities / sales |
| 43 | rotation receivables + inventory turnover in days |
| 44 | (receivables * 365) / sales |
| 45 | net profit / inventory |
| 46 | (current assets - inventory) / short-term liabilities |
| 47 | (inventory * 365) / cost of products sold |
| 48 | EBITDA (profit on operating activities - depreciation) / total assets |
| 49 | EBITDA (profit on operating activities - depreciation) / sales |
| 50 | current assets / total liabilities |
| 51 | short-term liabilities / total assets |
| 52 | (short-term liabilities * 365) / cost of products sold) |
| 53 | equity / fixed assets |
| 54 | constant capital / fixed assets |
| 55 | working capital |
| 56 | (sales - cost of products sold) / sales |
| 57 | (current assets - inventory - short-term liabilities) / (sales - gross profit – depreciation) |
| 58 | total costs /total sales |
| 59 | long-term liabilities / equity |
| 60 | sales / inventory |
| 61 | sales / receivables |
| 62 | (short-term liabilities *365) / sales |
| 63 | sales / short-term liabilities |
| 64 | sales / fixed assets |